\documentclass[12pt,a4paper]{article}
\usepackage[cp850]{inputenc}
\usepackage{amssymb}
\usepackage{amsmath,mathrsfs}
\usepackage{latexsym}
\usepackage{makeidx} 
\makeindex
\usepackage[dvips]{epsfig}
\usepackage{pstricks,pst-node}
\renewcommand{\baselinestretch}{1.5}
\newcommand{\eqa}{\begin{eqnarray}}
\newcommand{\neqa}{\end{eqnarray}}
\newcommand{\equ}{\begin{equation}}
\newcommand{\nequ}{\end{equation}}

\usepackage{bbm}

\begin{document}
\begin{titlepage} 
\renewcommand{\baselinestretch}{1}
\small\normalsize

\vspace{0.6cm}

\begin{center}   

{\LARGE \sc Fractal Quantum Space\hspace{0.15cm}-Time}

\vspace{2cm}
{\large Leonardo Modesto}\\

\vspace{1.7cm}
\noindent
{Perimeter Institute for Theoretical Physics, \\
31 Caroline St. N., Waterloo, ON N2L 2Y5, Canada}\\[12pt]
\end{center}   
 
\vspace*{1.6cm}

\begin{abstract}
In this paper we calculated the spectral dimension of loop quantum gravity (LQG) 
using the scaling property of the area operator spectrum on spin-network states and 
using the scaling property of the volume and length operators on Gaussian states. 
We obtained that the spectral dimension of the spatial section runs from $1.5$ to $3$, 
and under particular assumptions from $2$ to $3$ across a $1.5$ phase
when the energy of a probe scalar field decreases from high to low energy
in a fictitious time $T$. We calculated 
also the spectral dimension of space-time using the scaling of 
the area spectrum operator calculated on spin-foam models.
The main result is that the effective dimension is $2$ 
at the Planck scale and $4$ at low energy. This result is consistent with two 
other approaches to non perturbative quantum gravity: 
{\em causal dynamical triangulation} 
and {\em asymptotically safe quantum gravity}.
We studied the scaling properties of all the possible curvature invariants 
and we have shown that the singularity problem seems to be solved in the covariant formulation
of quantum gravity in terms of spin-foam models. For a particular form of the scaling
(or for a particular area operator spectrum) 
all the curvature invariants are 
regular also in the Trans-Planckian regime.

\end{abstract}

%
%
\end{titlepage}

\newpage
\tableofcontents
%
\newpage
\section{Introduction} 
\renewcommand{\theequation}{1.\arabic{equation}}
\setcounter{equation}{0}
\label{3intro}
In past years many approaches to quantum gravity
studied the fractal properties of the space-time.
In particular in {\em causal dynamical triangulation} (CDT) \cite{CDT} 
and {\em asymptotically safe quantum gravity} (ASQG) \cite{ASQG}, a fractal analysis 
of the space-time gives a two dimensional effective manifold
at high energy. In both approaches the spectral
dimension is ${\mathcal D}_s=2$ at small scales and ${\mathcal D}_s=4$ at large scales.
Recently the previous ideas have been applied in the context of {\em non commutativity} 
to a quantum sphere and
$\kappa$-Minkowski \cite{Dario} and in Loop Quantum Gravity \cite{Modesto0}.
The spectral dimension has been studied also in the cosmology of a Lifshitz universe
\cite{calcagni}.
Spectral analysis is a useful tool to understand the {\em effective form}
 of the space at small and large scales. 
We think that the fractal analysis could be also a useful tool to predict the behavior 
of the $2$-point and $n$-point functions at small scales and to attack the 
singularity problems of general relativity in a full theory
of quantum gravity. 

In this paper, we apply to {\em loop quantum gravity}
(LQG) \cite{book1} \cite{book2} the analysis 
developed in the context of ASQG 
by O. Lauscher and M. Reuter \cite{ASQG}.
In the context of LQG, we consider a spatial section, which is
a $3d$ manifold, and we extract the energy scaling of the 
metric in two different way from the area spectrum 
on the spin-network states and from the volume and length 
operators spectrum on Gaussian states. The result is the same 
until the Planck scale.
We apply the same analysis to the space-time using the area spectrum 
that is suggested by the spin-foam models \cite{DO}. 
In the space-time case, the result will be consistent with the spectral 
dimension calculated in the different approach of {\em non-perturbative
quantum gravity} \cite{CDT}, \cite{ASQG}.

In LQG, the average metric $\langle g_{\mu\nu}\rangle_{\ell}$
defines an infinite set of metric at different scales labeled by $\ell$. 
%
%
The metric average is over spin-network states, 
 $\langle g_{\mu\nu}\rangle_{\ell} := \langle s_{\ell}| g_{\mu\nu}| s_{\ell} \rangle$,
 where $| s_{\ell} \rangle = | \Gamma, j_e, \iota_v \rangle$ and $\ell^2= l_P^2 j$
is a diffeomorfism invariant length scale because $j$ is the $SU(2)$ 
Diff-invariant representation (in the paper we will consider also the average 
over Gaussian states obtaining the same result until the Planck scale).
The length $\ell$ is typically of the order of $1/k$, where $k$ is the momentum of a
probe field which plays the rule of microscope. 
The metrics $\langle g_{\mu\nu}\rangle_{k}$ 
one for each scale $k$ refer to the same physical system, the ``quantum spacetime'', but
describe its effective metric structure on different scales. An observer using
a ``microscope'' with a resolution $\ell \approx k^{-1}$ will perceive the universe
to be a Riemannian manifold with metric $\langle g_{\mu\nu}\rangle_k$. 
We suppose at every fixed $k$, $\langle g_{\mu\nu}\rangle_k$ is a smooth classical metric. 
But since the quantum spacetime is characterized by an infinity number of 
metrics $\{\langle g_{\mu\nu}\rangle_k \, , \,\, k=0, \dots +\infty\}$,
it can acquire very nonclassical and in particular
fractal features.

In a somewhat simplified form, the construction of a quantum spacetime within
LQG can be summarized as follows. We start from the Hilbert space of LQG
and we calculate the expectation value of the metric operator at any scale $\ell$, 
or for any $SU(2)$ representation $j$.
The quantum space-time 
is specified by the infinity of Riemannian metrics $\{\langle g_{\mu\nu}\rangle_j \,\, \big|
j=0,\cdots,+\infty\} \approx \{\langle g_{\mu\nu}\rangle_{\ell} \,\, \big| \,\, \ell =0,\cdots,+\infty\}
\approx \{\langle g_{\mu\nu}\rangle_k \,\, \big|
k=0,\cdots,+\infty\}$. 
An observer exploring the structure of the space-time using a microscope of resolution 
$\ell \approx 1/k$ ($k$ is the energy scale) will perceive the universe as a Riemannian manifold with the metric $\langle g_{ab} \rangle_k$ which is a fixed 
metric at every fixed scale $k$, the quantum space-time
 can have fractal properties because on different scales different metrics apply. 
In this sense the metric structure on the quantum space-time is given by an infinite set 
$\{ \langle g_{\mu \nu} \rangle_k; 0 \leqslant k < +\infty\}$ of ordinary metrics labelled by $k$
or by the Diff-invariant length scale $\ell :=l_P^2 j$.
In our analysis we will consider the expectation value $\langle g_{\mu \nu} \rangle_{k}$ 
as a smooth Riemannian metric because
we can approximate any metric with a weave state which is 
a spin network state with a large number of links and nodes 
that reweave the space. 
Microscopically it is a Planck size lattice but, at macroscopic scale, it appears as a continuum 
smooth metric.
Since we are interested to the fractal 
properties of the space (space-time) at different scales 
we suppose 
equal all
the representations on the spin-network links that across the surface of a given tetrahedron in the 
dual triangulation (at a fixed scale).
For this reason it is sufficient to analyze the metric scaling using an individual link.
If $|W_{\ell} \rangle$ denotes a weave state at the scale $\ell$
(the scale $\ell$ is defined such that all the $SU(2)$ representations $j$ that across a given surface 
are equal)
and $| W_{\ell_0} \rangle$ a weave state at the scale $\ell_0$, we have 
\begin{eqnarray}
\frac{\langle W_{\ell} | \hat{g}_{\mu \nu} | W_{\ell} \rangle}{\langle W_{\ell_0} | \hat{g}_{\mu \nu} | W_{\ell_0} \rangle} = \frac{\langle s_{\ell} | \hat{g}_{\mu \nu} | s_{\ell} \rangle}{\langle s_{\ell_0} | \hat{g}_{\mu \nu} | s_{\ell_0} \rangle}.
\label{metricWs0}
\end{eqnarray}
On the right hand side of (\ref{metricWs0}) we have a single link spin-network state
at the scales $\ell$ and $\ell_0$.
We are rescaling together all the representations dual to a given triangulation 
and then we 
can consider 
a single face of a single tetrahedron. For this reason on the right hand side of (\ref{metricWs0}) 
we have 
one single link spin-network duals to one face of one tetrahedron.
In other words all the scaling properties are encoded in a single link graph if
we are interesting to the scaling property of the metric and in particular to the 
fractal structure of the space (space-time). In the paper we will study the fractal
properties of the spatial section of LQG also using the expectation value of the
volume operator and length operator on Gaussian states. Those states can be 
treated as 
semiclassical until the Planck scale and then are useful for our intent.

 The paper is organized as follows. 
 In the first section we extract the information about the scaling property of the $3d$ spatial section metric from the area spectrum of LQG and from the average of the volume operator on of the length
 operator on Gaussian states.
Using the area operator spectrum in the context of spin-foam models we obtain
the scaling properties of the metric in $4d$.
In the second section we give a detailed review of the spectral dimension in diffusion 
processes. In the third section, we calculate explicitly the spectral dimension of the 
spatial section in LQG and of the space-time dimension. 
In the fourth section we show that the curvature invariant can be upper bounded 
using their scaling properties.

\section{Metric Scaling from 
the Area Spectrum} 
\renewcommand{\theequation}{2.\arabic{equation}}
\setcounter{equation}{0}
\label{SA} 
In this section we extract the scaling property of the expectation value of
the metric operator from the area spectrum obtained in LQG and Spin-Foams 
models.
\subsection{Metric scaling in LQG} 
One of the strongest results of LQG is the quantization of
the area, volume and recently length operators \cite{LoopOld}
\cite{EB}.
In this section, we recall the area spectrum and we deduce 
that the energy scaling of the $3d$-metric of the spatial section.
For a spin-network, 
$|\gamma; j_e, \iota_n \rangle$, without edges and nodes on the surface ${\mathcal S}$ we 
 consider the area spectrum 
\begin{eqnarray}
\hat{A}_{\mathcal S} |\gamma; j_e, \iota_n \rangle = 
8 \pi \gamma G_N \hbar \sum_{p \bigcap S} \sqrt{j_p(j_p+1)} |\gamma; j_e, \iota_n \rangle,
\label{area}
\end{eqnarray}
where $j_p$ are the representations on the edges that cross the surface ${\mathcal S}$.
Using (\ref{area}), we can calculate the relation between the area operator average \cite{LoopOld} 
for two different states of two different $SU(2)$ representations, $j$ and $j_0$, 
\begin{eqnarray}
\langle \gamma; j| \hat{A} |\gamma; j \rangle = \frac{l_P^2  \sqrt{j(j+1)}}{ l_P^2  \sqrt{j_0(j_0+1)}} 
\langle \gamma_0; j_0| \hat{A} |\gamma_0; j_0 \rangle.
\label{areass}
\end{eqnarray}
We can introduce the length squared defined by $\ell^2=l_P^2 j$ and the infrared length 
squared $\ell^2_0=l_P^2 j_0$.
Using this definition, we obtain the scaling properties of the area eigenvalues.
If $\langle  \hat{A}\rangle_{\ell}$ is the area average at the scale $l$ and 
$\langle  \hat{A}\rangle_{\ell_0}$ is the area average at the scale $l_0$ 
(with $\ell \leqslant \ell_0$), then we obtain the scaling relation 
\begin{eqnarray}
\langle  \hat{A}\rangle_{\ell} = \left\{\left[\ell^2(\ell^2+l_P^2)]/[\ell_0^2(\ell^2_0+l_P^2)\right] \right\}^{\frac{1}{2}}
\langle  \hat{A} \rangle_{\ell_0}.
\label{areall}
\end{eqnarray}

 \begin{figure}
\vspace{0.15cm}
 \begin{center}
 \includegraphics[height=4cm]{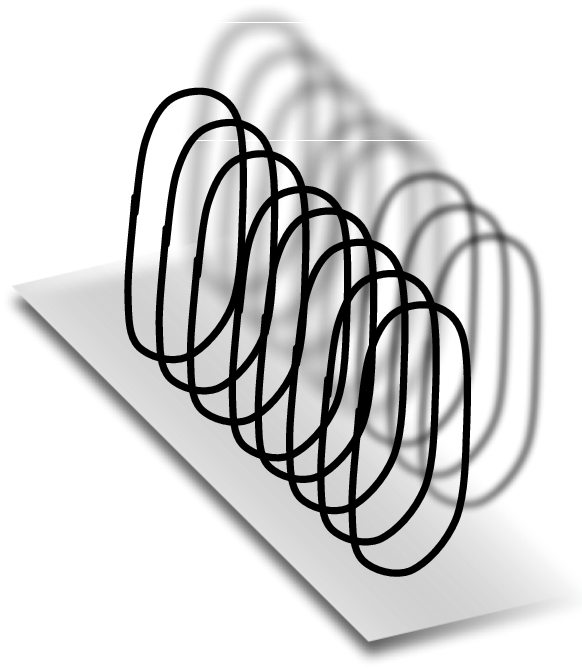}
 \hspace{0.5cm}
 \includegraphics[height=4cm]{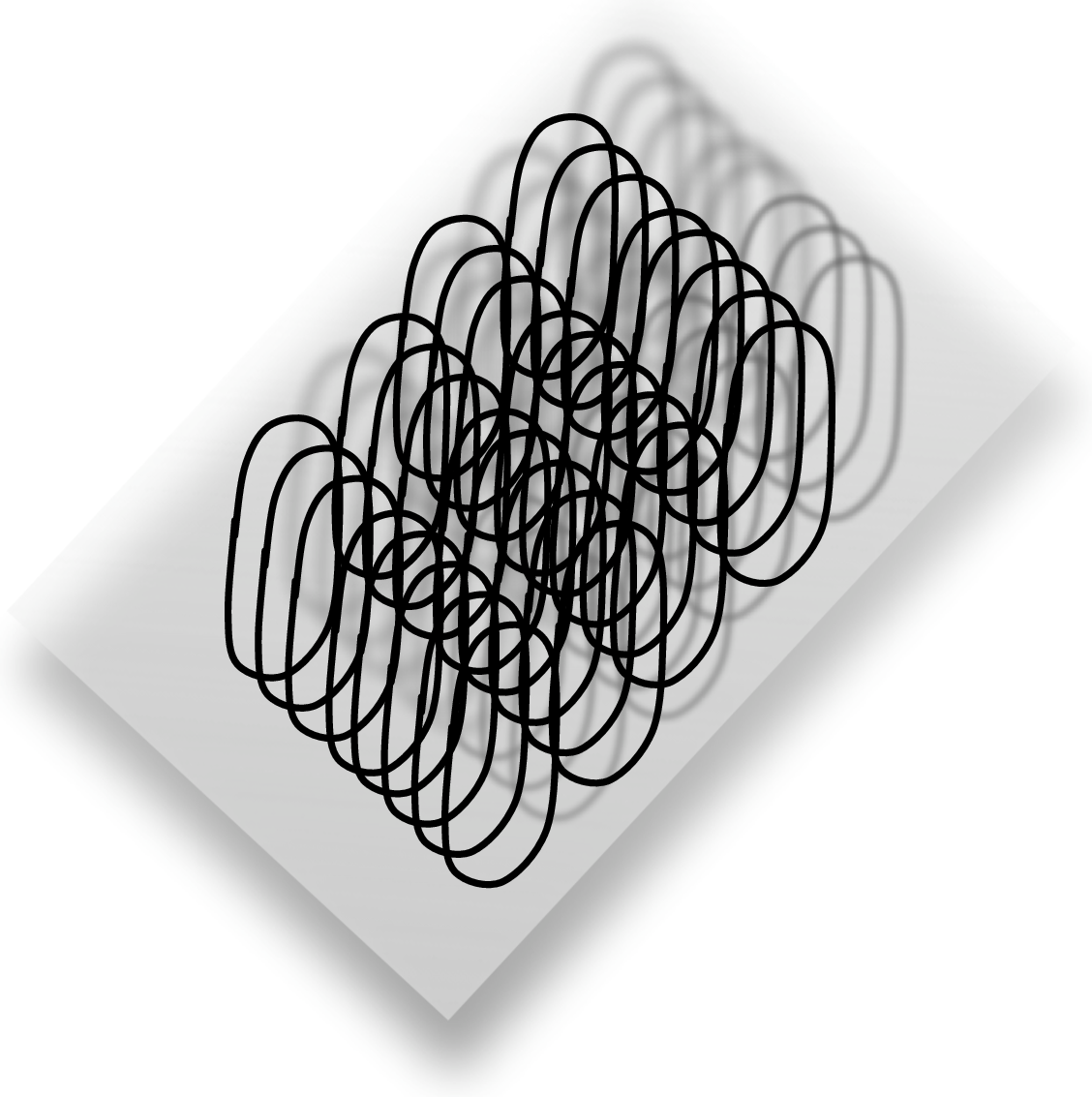}
 \hspace{1cm}
  \includegraphics[height=4cm]{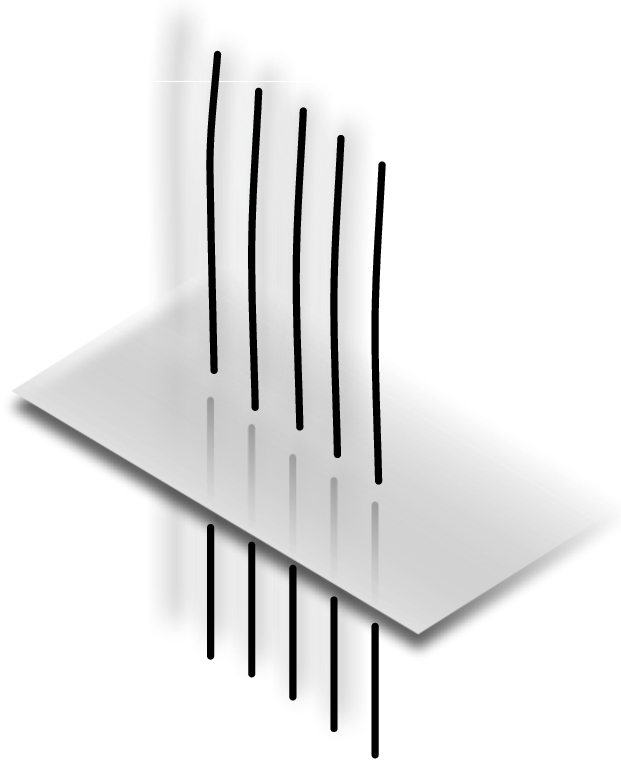}
    \end{center}
  \caption{\label{Plot04} 
  The first and the second pictures on the left 
represent two weave states with different 
loop's density. The picture represents only loops but argument is
valid for any spin-network. Any geometry can be approximate by
a weave state. The third picture represents the part of a spin-network 
that across a given surface.}
  \end{figure}
\begin{figure}
 \begin{center}
 \includegraphics[height=8.0cm]{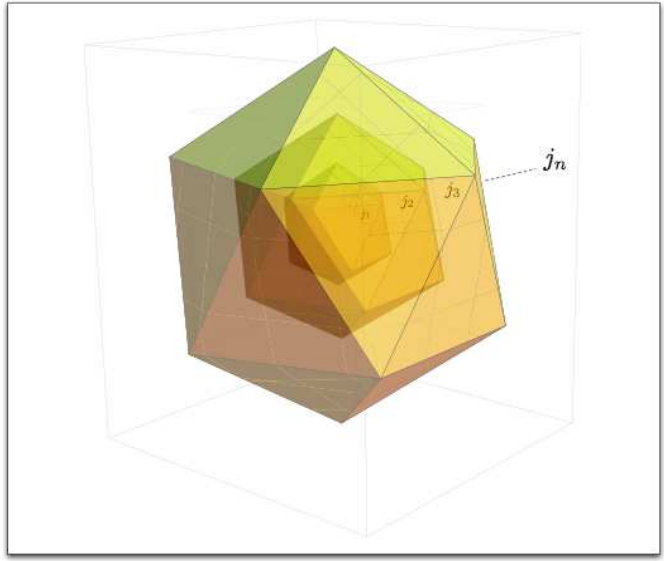}
    \end{center}
    \vspace{-0.3cm}
  \caption{\label{Scale0} 
  This picture represents the scaling of a 2-dimensional version of the 3-ball 
  explained in the text. We consider a simplicial decomposition of the 2-ball and we rescale the radius of the ball.
  A rescaling of the radius of the 2-sphere corresponds
  to a rescaling of all the representations $j$ dual to the triangle's area.  
 }
  \end{figure}
\begin{figure}
 \begin{center}
 \includegraphics[height=5.5cm]{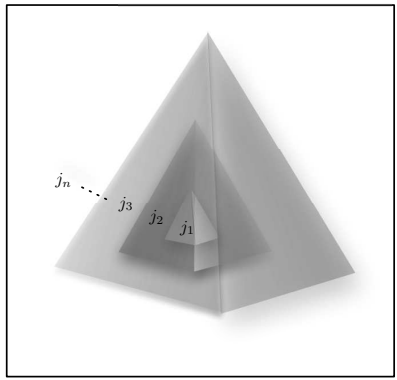}
    \end{center}
    \vspace{-0.3cm}
  \caption{\label{Scale} 
  Scaling of the tetrahedron for different values of the $SU(2)$ representation $j$
  or the length $\ell = l_P\sqrt{j}$. In the picture is represented one tetrahedron for 
  different values of the representation $j$. We can consider this tetrahedron and its scaling 
  as part of a simplicial decomposition. 
 }
  \end{figure}

We restricted our analysis to the case when a single edge crosses the surface because 
of the argument exposed at the end of the introduction and that we will go now to 
reconsider.  
In our analysis we will consider the expectation value $\langle g_{\mu \nu} \rangle_{k}$ 
of the metric at the energy scale $k$ 
as a smooth Riemannian metric because
we can approximate any metric with a weave state which is characterized  (for example) by a large 
number of loop 
that reweave the space. Since we are interested to the fractal 
properties of the space at different scales we consider equal all
the representations on the links that across a given surface at a fixed scale (see Fig.\ref{Plot04}).
We suppose to have a spin-network (weave state)
     that approximates the metric at a given fixed scale. We concentrate  
     our attention on a small region which is locally approximated by a single tetrahedron
     and we look at the scaling of the areas of such tetrahedron Fig.(\ref{Scale}).
     That scaling is given by the scaling of the areas of its faces
     because in our approximation we do not care about the non 
     commutativity of the metric. However, this approximation 
     is a good approximation if we look at the scaling and then at 
     the spectral properties of the space. Locally we suppose that any face of the tetrahedron
     is crossed by links with equal representations.
  In other words
     we can suppose to consider the spatial section as a 3-ball and triangulate it
     in a very fine way (the dual of the triangulation is a spin-network).
     Now we consider another 3-ball but at a smaller scale
     (also in this case the dual is another spin-network).
     Since we are considering the 3-ball at two different scales 
     all the representations of the spin-network states will be rescaled 
     of the same quantity (see Fig(\ref{Scale0})).
     If we concentrate on an individual tetrahedron of the 3-ball 
     triangulation we can extract its scaling considering just an area and 
     then just a dual reppresentation $j$
     (if all the representation of the weave state that cross that surface are equal).
     The representations involved in the spin-network will be different 
     to approximate a 3-ball 
      but the global scaling will be the same and this is what we will 
      use to calculate the spectral dimension.
We denote with $|W_{\ell} \rangle$ a weave state at the scale $\ell$
and with $|W_{\ell_0} \rangle$ a weave state at the scale $\ell_0$. 
We can think those weave states to describe the 3-ball at two different scales.
The scale is defined by $\ell := l_P \sqrt{j}$. 
All the $SU(2)$ representations $j$ that crosses a surface of a single 
tetrahedron 
 have the same value, as explained above.  
If $\ell$ and $\ell_0$ are two different scale and $N$ the number of links that across 
that surface we obtain 
\begin{eqnarray}
\frac{\langle W_{\ell} | \hat{A} | W_{\ell} \rangle}{\langle W_{\ell_0} | \hat{A} | W_{\ell_0} \rangle} = \frac{ N \, \langle s_{\ell} | \hat{A} | s_{\ell} \rangle}{ N \, \langle s_{\ell_0} | \hat{A} | s_{\ell_0} \rangle} =
\frac{\langle s_{\ell} | \hat{A} | s_{\ell} \rangle}{\langle s_{\ell_0} | \hat{A} | s_{\ell_0} \rangle},
\label{metricWs}
\end{eqnarray}
where $| s_{\ell_0} \rangle $, $| s_{\ell_0} \rangle $ are two spin-network such that only one
link of the graph crosses the surface we are considering; the spin-network are at the scale
$\ell$ and $\ell_{0}$ respectively. 
In other words all the scaling properties are encoded in a single link graph when
we are interesting to the scaling property of the metric and in particular to the 
fractal structure of the space. \\

The classical area operator can be related to the spatial metric $g_{ab}$ 
in the following way.
The classical area operator can be expressed in terms of the density triad 
operator,
$A_{\mathcal S}=\int_{\mathcal S} \sqrt{n_a E^{a}_i n_b E^{b}_{i}} d^2\sigma$,
and the density triad is related to the three dimensional triad by 
$e^{a}_i = E^{a}_i /\sqrt{{\rm det} E}$ and $\sqrt{{\rm det} E} = {\rm det} e$.
If we rescale the area operator by a factor ${\mathcal Q}^2$, 
$A \rightarrow A^{\prime} = {\mathcal Q}^2 A$, consequently 
the density triad scales by the same quantities, 
$E^a_i \rightarrow E^{a \prime}_i = {\mathcal Q}^2 E^a_i$.
The triad instead, using the above relation, scales as
$e^a_i \rightarrow e^{a \prime}_i = {\mathcal Q}^{-1} e^a_i$ 
and its inverse  
$e^i_a \rightarrow e^{i \prime}_a = {\mathcal Q} e^i_a$. The 
metric on the spatial section is related to the triad by 
$g_{ab} = e^i_a e^j_b \delta_{ij}$ and then it scales 
as $g_{ab} \rightarrow g_{ab}^{\prime} = {\mathcal Q}^2 g_{ab}$,
or, in other words, the metric scales as the area operator.
Using (\ref{areall}), 
we obtain the following scaling for the metric 
\begin{eqnarray}
\langle  \hat{g}_{ab} \rangle_{\ell} =  \left\{\left[\ell^2(\ell^2+l_P^2)]/[\ell_0^2(\ell^2_0+l_P^2)\right] \right\}^{\frac{1}{2}}
\langle  \hat{g}_{ab}\rangle_{\ell_0}.
\label{metricll}
\end{eqnarray}
The scaling (\ref{metricll}) is not an assumption if we restrict our attention 
to diagonal part of the metric (see the last part of this section); this assumption is justified
because we are not interested to the non commutativity of the metric at fixed scale
but instead to the the metric at different scales. 
We have a fixed manifold and also a fixed 
metric at any scale $\ell$. Formula (\ref{metricll}) provides 
a relation between two metrics at different scales $\ell$ and $\ell_0$.
If we want to explore the spatial section structure at a fixed length $\ell$ 
we should use a microscope of resolution $\ell$ or, in other words, 
we should use (for example) 
a probe scalar field of momentum $k \approx 1/\ell$ ($\Delta k \, \Delta \lambda \gtrsim 1$,
this approximation can be justified using Riemann normal coordinates in a small region
of the manifold. This approximation is related to the curvature of the manifold and
not to the scaling properties of the metric). 
The scaling property of the metric in terms of $k$ can be obtained 
by replacing: $\ell \approx 1/k$, $\ell_0\approx 1/k_0$ and $l_P\approx 1/E_P$,
where $k_0$ is an infrared energy cutoff and $E_P$ is the Planck energy.
The scaling
of the metric as function of $k$, $k_0$ and $E_P$ is, 
\begin{eqnarray}
\langle  \hat{g}_{ab} \rangle_k= [ k_0^4(k^2+E_P^2)/[ k^4(k_0^2+E_P^2)]]^{\frac{1}{2}}
\langle  \hat{g}_{ab}\rangle_{k_0}.
\label{metrickk}
\end{eqnarray}
In particular 
we will use the scaling properties of the inverse metric (see also below the last part of this section),
\begin{eqnarray}
\langle  \hat{g}^{ab} \rangle_k= \left[\frac{ k^4(k_0^2+E_P^2)}{ k^4_0(k^2+E_P^2)} \right]^{\frac{1}{2}}
\langle  \hat{g}^{ab}\rangle_{k_0}.
\label{metrickkINV}
\end{eqnarray}
We define the scaling factor in (\ref{metrickkINV}), introducing a function $F(k)$: $\langle  \hat{g}^{ab} \rangle_k=F(k)\langle  \hat{g}^{ab}\rangle_{k_0}$. 
From the explicit form of $F(k)$ we have three different phases where the behavior of $F(k)$ can be approximated as follows,
\begin{eqnarray}
F(k) \approx \left\{ \begin{array}{lll} 
         1 & , \, k \approx k_0, \\ 
         k^2  & , \, k_0 \ll k \ll E_P,,\\
         k  &  , \, k\gg E_P.
        \end{array} \right. 
\label{Flimits}
\end{eqnarray}
We consider $F(k)$ to be constant for $k\lesssim k_0$; in particular we require that $F(k)\approx 1$, 
$\forall k \lesssim k_0$. To simplify the calculations without modifying the scaling
properties of the metric, we introduce the new function 
${\mathcal F}(k) = F(k) +1$.
The behavior of ${\mathcal F}$ is exactly the same as in (\ref{Flimits}) but with better
properties in the infrared limit which one useful in the calculations. 
We define here the scale function ${\mathcal F}(k)$ for future reference in the next sections,
\begin{eqnarray}
{\mathcal F}(k)= \left[\frac{ k^4(k_0^2+E_P^2)}{ k^4_0 (k^2+E_P^2)} \right]^{\frac{1}{2}} +1.
\label{Fprime}
\end{eqnarray}

We can make more clear the argument of this section 
in the following way.
The metric is related to the density triad by $\sqrt{g} g^{ab}(x) = E^a_i E^b_i(x)$. 
If we take a tetrahedron as our chunk of space (substantially 
this correspond to take four valent spin-networks and to identify the point $x$ with the 
node $n$ dual to the tetrahedron) the metric can be expressed 
in terms of the area of the faces and the angles of the tetrahedron
\begin{eqnarray}
\sqrt{g} g^{ab}: = \frac{1}{(8 \pi \gamma l_P^2)^2} 
\left( \begin{array}{ccc} 
A_1^2 & A_1 A_2 \cos \theta_{12} & A_1 A_3 \cos \theta_{13}  \\ 
A_1 A_2 \cos \theta_{12}  & A_2^2 & A_2 A_3 \cos \theta_{23}  \\
A_1 A_3 \cos \theta_{13}  & A_2 A_3 \cos \theta_{23}  & A_3^2  \end{array} \right).
\label{metricA}
\end{eqnarray}
The area $A_i$ are three areas that shide a node and $\cos \theta_{ij}$ is the cosine 
of the angle between the normals to the face $i$ and $j$.
Because we are interested in the scaling of the metric 
we can consider an equilateral tetrahedron, $A_1 = A_2 = A_3 :=A$, and then all 
the $SU(2)$ representations $j$ of the dual spin-network are equal. 
For the same reason we do not quantize the $\cos \theta_{ij}$ operators 
because they are related to the quantum anisotropy or non commutativity 
of the metric that 
does not contain information about the scaling (see Fig.\ref{Scale}). 
Under those assumptions the operator $E^aE^b$ is diagonal 
on the spin-network states, because the angular part is frozen,
and reduces to 
 \begin{eqnarray}
\widehat{ E^a_i E^b_i} 
:=\frac{\hat{A}^2}{(8 \pi \gamma l_P^2)^2} 
\left( \begin{array}{ccc} 
1 & \cos \theta_{12} &    \cos \theta_{13} \\ 
\cos \theta_{12}  & 1 &   \cos \theta_{23} \\
\cos \theta_{13}  &  \cos \theta_{23}  & 1  \end{array} \right)
:= \frac{\hat{A}^2_j}{(8 \pi \gamma l_P^2)^2}  \, M^{ab}_{\theta}
\label{metricAe}
\end{eqnarray}
We indicate the spin-network with $|j \rangle$ and calculate the expectation 
value of  $\sqrt{g} g^{ab} = E^a_i E^b_i$ (\ref{metricA}),
\begin{eqnarray}
&& \langle j | \widehat{\sqrt{g} g^{ab}}| j \rangle = 
 \langle j | \widehat{ E^a_i E^b_i} | j \rangle 
= j(j+1) \,  \, M^{ab}_{\theta}.
\label{metricAe2}
\end{eqnarray}
Since under our assumption, dictated from the physics we want to study
it is simple to extract the determinant and obtain the spectrum of the inverse metric, we have:
\begin{eqnarray}
&& \langle j | \widehat{g^{ab}} | j \rangle=\langle j |
\widehat{E^a_i E^b_i} \, [{\rm det}  (\widehat{E^a_i E^b_i}  ) ]^{-1/2} | j \rangle = 
\frac{ j(j+1)   \, M^{ab}_{\theta} }{\sqrt{  
(j(j+1))^3  \, {\rm det} ( M^{ab}_{\theta}} )}.
\label{ginv}
\end{eqnarray}
The scaling of the metric is defined by looking on to $SU(2)$ representations $j$ and $j_0$
that define two different scales 
and calculating the following ratio, 
\begin{eqnarray}
\frac{\langle j | \widehat{g^{ab}} | j \rangle}{\langle j_0 | \widehat{g^{ab}} | j_0 \rangle} =
\frac{\sqrt{ \langle j_0 | \hat{A^2} | j_0 \rangle}}{\sqrt{ \langle j | \hat{A^2} | j \rangle}}.
\label{ginvs}
\end{eqnarray}
We stress that the scaling is independent from the angular variables because we are interested
in metrics at different scales and we do not take care of the different directions at a fixed scale.

\subsection{Metric Scaling in Spin-Foams}\label{SFS}
  We can repeat the scaling analysis above in the case of a four dimensional
  spin-foam model. 
  In the spin-foam models framework the starting point is 
  a simplicial decomposition of the space-time in $4$-simplexes.
  Any simplex is made of $5$ tetrahedron and we can consider the area 
  operator associated with whatever face of a general tetrahedron of the
  decomposition. The face can be directed in any direction and then can be 
  space-like or time-like.
  The area operator commutes with all the 
  constraints and then is a good observable. 
  The result useful for our aim is that in the context of 
  spin-foams models we can have three possible area spectrum:
  $A_j=l_P^2 j$, $A_j=l_P^2 (2j+1)$ and $A_j=l_P^2 \sqrt{j(j+1)}$.
   In the first case, when the area eigenvalues are $A_j=2 l_P^2 j$ \cite{DO},
   the scaling of the $4d$ metric is 
  \begin{eqnarray}
\langle  \hat{g}^{\mu \nu} \rangle_k= \frac{ k^2}{ k^2_0}
\langle  \hat{g}^{\mu \nu}\rangle_{k_0},
\label{metrickkINV4d}
\end{eqnarray}
where $\mu, \nu =1, \dots,4$. Given the explicit form of the scaling 
in (\ref{metrickkINV4d}), 
we introduce the new scaling function, 
\begin{eqnarray}
{\mathbb S}_1(k) = \frac{k^2}{k_0^2} +1 \,\,\,\,\,\,\,(A_j= 2 l_P^2 j).
\label{S1}
\end{eqnarray} 
The infrared modification, introduced by hand,  
does not change the high energy behavior of the scaling function
and we can take $k \in [0, +\infty[$ in the calculations.
A different ordering in the area operator quantization can give 
a different spectrum $A_j=l_P^2  (2j+1)$ \cite{DO}, \cite{DO2}. The scaling function 
in this case is 
\begin{eqnarray}
{\mathbb S}_2 = \frac{k^2 (k_0^2+ 2 E_p^2)}{k_0^2 (k^2+2 E_p^2)} +1 \,\,\,\,\,\,\, (A_j=l_P^2  (2j+1)),
\label{scaling2}
\end{eqnarray} 
where we have introduced the usual infrared modification: $+1$.
We can consider also in $4D$ the same scaling of the $3D$ spatial section,
this corresponds to the matching of the area spectrum that comes 
from the spin-foam model with the kinematical area spectrum of LQG.
The result is (\ref{Fprime}) specialized to four dimension,
\begin{eqnarray}
{\mathbb S}_3(k)= \left[\frac{k^4(k_0^2+E_P^2)}{k^4_0 (k^2+E_P^2)} \right]^{\frac{1}{2}} +1
 \,\,\,\,\,\,\, (A_j=l_P^2 \sqrt{j(j+1)}).
\label{scaling3}
\end{eqnarray}

\section{Metric Scaling in LQG from Gaussian States} 
\begin{figure}
 \begin{center}
 \hspace{0.0cm}
 \includegraphics[height=6.0cm]{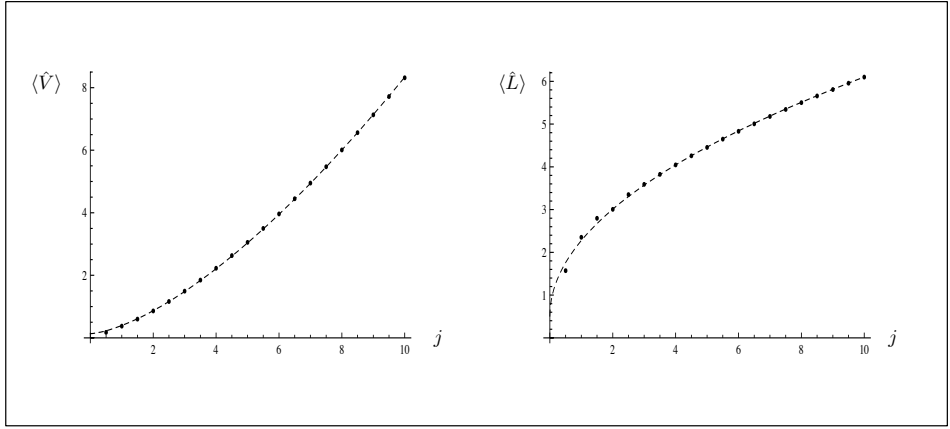}
     \end{center}
    \vspace{-0.3cm}
  \caption{\label{VLsemi} 
 The two plots represent respectively the expectation value 
 of the volume and length operators on Gaussian states.
 The expectation value of the volume is well fitted with 
 the dashed line by the function $j^{3/2}$ and the length operator expectation 
 value by the function $j^{1/2}$ it is relevant to observe
 that the perfect matching until Planck scales (in the plots $j \approx 1$).
 }
  \end{figure}

\renewcommand{\theequation}{3.\arabic{equation}}
\setcounter{equation}{0}
\label{SVL}
In this section we extract the scaling of the metric using a recent
result of Bianchi \cite{EB}.
In \cite{EB} has been calculated the expectation value of the volume operator and 
in particular of the length operator on a gaussian state.
The author consider a
$4$-valent monochromatic spin-network (the valence of the node is four and
all the $SU(2)$ representations are equal: $j_1 =j_2 = j_3 =j_4 :=j$) and the state
introduced by Rovelli and Speziale \cite{RS},
\begin{eqnarray}
| c \rangle = \sum_i \frac{3^{1/8}}{(\pi i_0)^{1/4}} 
e^{- \frac{\sqrt{3}}{2} \frac{(i - i_0)^2}{i_0}} e^{i \phi_0 i} |i_{12} \rangle,
\label{gaussian}
\end{eqnarray}
where $i_0 = 2 j/\sqrt{3}$, $\phi_0 = \pi/2$ and $|i_{12} \rangle$ is the basis state  
associated to the intertwining tensor for the representations $j_1$, $j_2$ that are 
equal in our particular case. 
This state has good semiclassical geometric properties and the interested 
reader is invited to consult the original 
paper for the details \cite{RS}.
The expectation value of the volume and the length operator on (\ref{gaussian}) is
\begin{eqnarray}
&& \langle c | \hat{V} | c \rangle \approx l_P^3 \, j^{3/2}, \nonumber \\
&& \langle c | \hat{L} | c \rangle \approx l_P \, j^{1/2}.
\label{VLcoh}
\end{eqnarray}
This behavior is correct also for small values of $j$ as it is evident from the plots
in Fig.\ref{VLsemi}.
There is a new version of Gaussian states that confirm this result for any value 
of the representation $j$.
 The choice of monochromatic representation is not restrictive but it is necessity 
 because we are interesting to the scaling property of the metric at different scales 
 and the scale is defined by $\ell = l_P j^{1/2}$.
The results in (\ref{VLcoh}) suggests the following scale of the metric, 
of the metric,
\begin{eqnarray} 
\langle c, j | \hat{g}_{ab} | c, j  \rangle = 
\frac{j}{j_0} \langle c, j_0 | \hat{g}_{ab} | c, j_0  \rangle \,\,\,\,\,  \rightarrow \,\,\,\,\, 
 \langle c_{\ell} | \hat{g}_{ab} | c_{\ell}  \rangle = 
\frac{\ell^2}{\ell^2_0} \langle c_{\ell_0} | \hat{g}_{ab} | c_{\ell_{0}} \rangle 
\label{VLcoh2}
\end{eqnarray}
where $j$ and $j_0$ are two different representation that satisfy 
the relation $j > j_0$ and $\ell = l_P \sqrt{j}$, $\ell_0 = l_P \sqrt{j_0}$.
The scaling of the metric for a test field of momentum $k$ is 
\begin{eqnarray} 
 \langle c_{k} | \hat{g}_{ab} | c_{k}  \rangle = 
\frac{k_0^2}{k^2} \langle c_{k_0} | \hat{g}_{ab} | c_{k_{0}} \rangle.
\label{VLcoh3}
\end{eqnarray}
This result coincides with the scaling obtained on spin-network states for $k \lesssim E_P$.

\section{The spectral dimension}
\renewcommand{\theequation}{4.\arabic{equation}}
\setcounter{equation}{0}
\label{Sec3}
In this section we determine the spectral dimension ${\cal D}_{\rm s}$ 
of the quantum space and the quantum space-time.
This particular definition of a fractal dimension is borrowed
from the theory of diffusion processes on fractals \cite{avra} and is easily 
adapted to the
quantum gravity context. 

Let us study the diffusion of a scalar test (probe) particle on a $d$-dimensional 
classical Euclidean manifold with a fixed smooth metric $g_{\mu\nu}(x)$.
The corresponding heat-kernel $K_g(x,x';T)$ giving the probability for the
particle to diffuse from $x'$ to $x$ during the fictitious diffusion time $T$
satisfies the heat equation
\begin{eqnarray}
\label{heateq}
\partial_T K_g(x,x';T)=\Delta_g K_g(x,x';T)
\end{eqnarray}
where $\Delta_g$ denotes the scalar Laplacian: 
$\Delta_g\phi\equiv g^{-1/2}\,\partial_\mu(g^{1/2}\,g^{\mu\nu}\,\partial_\nu
\phi)$. The heat-kernel is a matrix element of the operator 
$\exp(T\,\Delta_g)$, 
\begin{eqnarray}
K_g(x,x';T) =\langle x^{\prime} | \exp(T\,\Delta_g) | x \rangle.
\end{eqnarray} 
In the random walk picture its trace per unit volume,
\begin{eqnarray}
\label{trace}
P_g(T)&\equiv& V^{-1}\int d^dx\,\sqrt{g(x)}\,K_g(x,x;T)\,\;\equiv\,\; 
V^{-1}\,{\rm Tr}\,
\exp(T\,\Delta_g)\;,
\end{eqnarray}
has the interpretation of an average return probability. (Here $V\equiv\int
d^dx\,\sqrt{g}$ denotes the total volume.) It is well known that $P_g$
possesses an asymptotic expansion (for $T\rightarrow 0$) of the form
$P_g(T)=(4\pi T)^{-d/2}\sum_{n=0}^\infty A_n\,T^n$. For an infinite flat
space, for instance, it reads $P_g(T)=(4\pi T)^{-d/2}$ for all $T$. Thus,
from the knowledge of the function $P_g$, one can recover the dimensionality of the target
manifold from the $T$-independent logarithmic derivative
\begin{eqnarray}
\label{dimform}
d=-2\frac{d\ln P_g(T)}{d\ln T}
\end{eqnarray}
This formula can also be used for curved spaces and spaces with finite volume
$V$ provided $T$ is not taken too large. 

In quantum gravity 
it is natural to replace $P_g(T)$ by its expectation value on the spin-network states
: $| s \rangle$ (we will calculate also the spectral dimension of the spatial section on 
Gaussian states and in that case the expectation value is over the state $|c \rangle$ 
of section (\ref{SVL})). 
Symbolically,
\begin{eqnarray}
\label{pexpect}
P(T):= \langle \hat{P_{g}}(T)\rangle = \langle s | \hat{P_{g}}(T) | s \rangle
\approx  P_{\langle s | \hat{{g}}| s \rangle}(T) = P_{\langle  g  \rangle}(T) .
\end{eqnarray}
The third relation is not  an equality but an approximation because it is valid only in the case
that the metric operator is diagonal on the state considered.  
This is not true in general on spin-network states 
but, in the case we are interested in the scaling properties of the metric, 
we will not consider the non diagonal terms that are related to the non commutativity  
(or angular part) of the metric. In particular we do not quantize the angular part 
of the metric. 
This assumption justifies (\ref{pexpect}). 
On the other part on gaussian states (\ref{pexpect}) is a good approximation in the
large $j$ limit. 
%
For the space-time spectral dimension the scalar product 
in (\ref{pexpect}) is the physical scalar product that defines the dynamics.
In our context the dynamics is defined by the spinfoam models \cite{DO}.

Given $P(T)$, the spectral dimension of the quantum space or 
space-time is defined in analogy with (\ref{dimform}):
\begin{eqnarray}
\label{specdim}
{\cal D}_{\rm s}=-2\frac{d\ln P(T)}{d\ln T}.
\end{eqnarray}
%

The fictitious diffusion process takes place on a ``manifold''
which, at every fixed scale $\ell \approx 1/k$, is described by a smooth Riemannian metric
$\big<g_{\mu\nu}\big>_k$. While the situation appears to be classical at fixed
$k$, nonclassical features emerge 
since at different scales different metrics apply. 
The metric depends on the scale at which the spacetime structure is probed
by a fictitious scalar field.

In quantum geometry the equation (\ref{heateq}) is replaced with 
the expectation value on the spin-network states,
\begin{eqnarray}
\label{heateqs}
 \partial_T \langle s|K_{\hat{g}}|s \rangle = 
\langle s| \Delta_{\hat{g}} K_{\hat{g}}| s \rangle \,\,\, \approx \,\,\,  
 \partial_T K_{\langle s |\hat{g}| s \rangle } = 
\Delta_{\langle s |\hat{g}| s \rangle } K_{\langle s | \hat{g} | s \rangle } \, ,
\end{eqnarray}
where $K_{\dots} = K_{\dots}(x,x';T)$. 
We denote the scaling of the metric operator by a general function that we 
will specify case by case farther on in the paper,
\begin{eqnarray}
\langle s | \hat{g}^{\mu \nu} | s \rangle =
{\mathbb S}(\ell, \ell_0) \langle s_0 | \hat{g}^{\mu \nu} | s_0 \rangle   
\,\,\, \rightarrow \,\,\, \langle \hat{g}^{\mu \nu} \rangle_k =\mathbb{S}_k (k,k_0)\,
\langle \hat{g}^{\mu \nu} \rangle_{k_0}.
\end{eqnarray}
where we have shorten, $|s_{\ell }\rangle := | s \rangle$ and $|s_{\ell_0}\rangle := | s_0 \rangle$.

The nonclassical features are encoded in the properties of the diffusion
operator. We define the covariant Laplacians corresponding to the 
metrics $\big<g_{\mu\nu}\big>_k$ and $\big<g_{\mu\nu}\big>_{k_0}$ by
$\Delta(k)$ and $\Delta(k_0)$, respectively at the scale $k$ and $k_0$. 
%
We extract the scaling of the Laplacian operators from the behavior of the
metric at different scales 
\begin{eqnarray}
\label{opscale0}
\Delta_{\langle s |\hat{g}| s \rangle } 
=\mathbb{S}_{\ell}(\ell ,\ell_0)\,\Delta_{\langle s_0 |\hat{g}| s_0 \rangle } \,\,\, \rightarrow \,\,\, \Delta(k) =\mathbb{S}_k (k,k_0)\,\Delta(k_0).
\end{eqnarray}
We suppose the diffusion process  involves (approximately) only a small interval of 
scales near $k$ 
over which the expectation value of the metric does not change much 
then the corresponding heat equation contains the $\Delta(k)$ for this specific, fixed
value of the momentum scale $k$:
\begin{eqnarray}
\label{heateqk}
\partial_T K(x,x';T)&=&\Delta(k) K(x,x';T), 
\end{eqnarray}
The equation (\ref{heateqk}) is exactly (\ref{heateqs}) where we suppressed 
the index $\langle s |\hat{g}| s \rangle$, and introduced the Laplacian at the scale $k$ in terms 
of the Laplacian at the scale $k_0$.

\noindent Denoting the eigenvalues of $-\Delta(k_0)$ by ${E}_n$
and the corresponding eigenfunctions by $\phi_n(x) = \langle x| E_n \rangle$, 
we have the following eigenvalue equation for the Laplacian 
\begin{eqnarray}
\Delta(k_0) |E_n \rangle = - E_n |E_n \rangle.
\label{Deltak0}
\end{eqnarray}
Using (\ref{Deltak0}) the equation (\ref{heateqk}) is solved by
\begin{eqnarray}
K(x,x';T)&=&\sum\limits_n\phi_n(x)\,\phi^*_n(x')\,\exp\big(-{\mathbb S}_k(k, k_0)\, E_n
\,T\big).
\label{kernexp}
\end{eqnarray}
%
%
{\bf Proof 1 of (\ref{kernexp}).}
We want to obtain 
$K_{\langle g \rangle_k}(x,x';T) := K(x,x';T)$ using the definition given at the beginning 
of this section. Using (\ref{opscale0}) and (\ref{Deltak0}) the solution of (\ref{heateqk}) is: 
\begin{eqnarray}
&& K_{\langle g \rangle_k}(x,x';T) 
= \langle x^{\prime}| e^{T\Delta_{\langle g \rangle_k}} | x \rangle \nonumber \\
&& \hspace{0.0cm} = \sum_n \sum_{n^{\prime}} \langle x^{\prime}| E_{n^{\prime}} \rangle \langle E_{n^{\prime}}| 
e^{T\Delta_{\langle g \rangle_k}} | E_{n} \rangle \langle E_{n}| x \rangle \nonumber \\
&& \hspace{0.0cm} = \sum_n \sum_m \phi^*_{n^{\prime}}(x^{\prime}) \langle E_{n^{\prime}}| 
e^{T \, {\mathbb S}_k (k, k_0) \Delta(k_0)} | E_{n} \rangle \phi_{n}(x) \nonumber \\
&& = \sum_n \sum_{n^{\prime}} \phi^*_{n^{\prime}}(x^{\prime}) \langle E_{n^{\prime}}| 
e^{ - T \, {\mathbb S}_k (k, k_0) E_n} | E_{n} \rangle \phi_{n}(x) \nonumber \\
&& = \sum_n \sum_{n^{\prime}} \phi^*_{n^{\prime}}(x^{\prime}) \phi_{n}(x) \, \delta_{n^{\prime}, n} \,  
e^{- T \, { \mathbb S}_k (k, k_0) E_n}\nonumber \\
&& =\sum_n  \phi^*_n(x^{\prime}) \phi_{n}(x) \, 
e^{- T \, { \mathbb S}_k (k, k_0) E_n}.
\label{Ktra}
\end{eqnarray}
 %
%
%
{\bf Proof 2 of (\ref{kernexp}).} 
 We show below that the left hand side and the right hand side of (\ref{heateqk}) 
are equal.
\begin{eqnarray}
&& {\rm LHS} \, : \,\,\, \partial_T K(x,x';T) = \sum\limits_n\phi_n(x)\, \phi^*_n(x')
(-{\mathbb S}_k(k, k_0)\, E_n )
\,\exp\big(-{\mathbb S}_k(k, k_0)\, E_n
\,T\big), \nonumber 
\end{eqnarray}
\begin{eqnarray}
&& {\rm RHS} \, : \,\,\, \bigtriangleup_x(k) K(x,x';T) = \sum\limits_n (\Delta_x(k) \phi_n(x)) \, \phi^*_n(x')
\,\exp\big(-{\mathbb S}_k(k, k_0)\, E_n
\,T\big) \nonumber \\
&& \hspace{4.2cm} = \sum\limits_n (-{\mathbb S}_k(k, k_0)\, E_n  \phi_n(x)) \, \phi^*_n(x')
\,\exp\big(-{\mathbb S}_k(k, k_0)\, E_n
\,T\big). \nonumber
\label{L}
\end{eqnarray} 
From the knowledge of the propagation kernel (\ref{Ktra}) we can time-evolve any
initial probability distribution $p(x;0)$ according to
$p(x;T)=\int d^4x'\,\sqrt{g_0(x')}\,K(x,x';T)\,p(x';0)$, where $g_0$ is the 
determinant of $\big<g_{\mu\nu}\big>_{k_0}$. If the initial distribution has 
an eigenfunction expansion of the form $p(x;0)=\sum_n C_n\,\phi_n(x)$ we
obtain, 
\begin{eqnarray}
 p(x;T)= \sum_n C_n\,\phi_n(x)\, e^{-{\mathbb S}(k, k_0)\,{E}_n\,T}. 
 \label{probexp0}
\end{eqnarray}
%
%
{\bf Proof of (\ref{probexp0}).}
\begin{eqnarray}
\label{probexp}
&& \hspace{-1.37cm} p(x;T)= \int d^4x'\,\sqrt{g_0(x')}\,K(x,x';T)\,p(x';0) = \nonumber \\
&& = \sum_n \sum_m \, \int d^4x'\,\sqrt{g_0(x')}\,
  \phi^*_n(x^{\prime}) \phi_{n}(x) \, 
e^{- T \, { \mathbb S}_k (k, k_0) E_n}  \,
 C_m\,\phi_m(x^{\prime})
\nonumber  \\
&& =\sum_n C_n\,\phi_n(x)\, e^{-{\mathbb S} (k, k_0)\, E_n\,T}
\end{eqnarray}
%
From second to third line we used the weave function normalization property:
\begin{eqnarray}
 \langle E_n| E_m \rangle = \int d^4x'\,\sqrt{g_0(x')} \, \phi^*_n(x^{\prime}) \phi_{n}(x^{\prime}) = 
 \delta_{n, m}.
 \label{norm}
 \end{eqnarray}
If the $C_n$'s are significantly different from zero only for a single
eigenvalue ${E}_n$, we are dealing with a single-scale problem and 
then we can identify 
$k^2={E}_n$. 
However, in general the $C_n$'s are different from zero over a wide range of
eigenvalues. In this case we face a multiscale problem where different modes
$\phi_n$ probe the spacetime on different length scales.

If $\Delta(k_0)$ is the Laplacian on the 
flat space, the eigenfunctions $\phi_n
\equiv\phi_p$ are plane waves with momentum $p^\mu$, and they probe 
structures on a length scale $\ell$ of order $1/|p|$. Hence, in terms of the
eigenvalue $E_n\equiv{ E}_p=p^2$ the resolution is $\ell \approx
1/\sqrt{{E}_n}$. This suggests that when the manifold is probed by a
mode with eigenvalue ${E}_n$ it ``sees'' the metric 
$\big<g_{\mu\nu}\big>_k$ for the scale $k=\sqrt{{E}_n}$. Actually the
identification $k=\sqrt{{E}_n}$ is correct also for a curved space
because the parameter $k$ just identifies the scale we are probing.
Therefore we can conclude that under the spectral sum of (\ref{probexp}) we must
use the scale $k^2={E}_n$ which depends explicitly on the resolving power
of the corresponding mode. 
In eq. (\ref{kernexp}), ${\mathbb S}(k, k_0)$ can be
interpreted as ${\mathbb S}({E}_n)$. 
Thus we obtain the traced propagation kernel
\begin{eqnarray}
\label{trpropk}
P(T)=V^{-1}\;\sum_n e^{-{\mathbb S}({ E}_n)\,{E}_n\,T} 
= V^{-1}\;{\rm Tr}\, \left(e^{ {\mathbb S}(-\Delta(k_0) )\,\Delta(k_0)\,T}\right).
\end{eqnarray}
It is convenient to choose $k_0$ as a macroscopic scale in a regime where there are not strong  quantum gravity effect. \\
{\bf Proof of (\ref{trpropk})}.
\begin{eqnarray}
&& \hspace{0.1cm} P(T) =  \langle s | \left(V^{-1} \, \int \sqrt{g} \, d^d x \, K_{g}(x, x;T) \right)| s \rangle \nonumber \\
 && \hspace{1.2cm} 
 = \langle s | \left( V^{-1} \, \int \sqrt{g} \, d^d x \, 
\langle x | e^{T \Delta_g} | x \rangle \right) | s \rangle \nonumber \\
&& \hspace{1.2cm} \approx   V^{-1}(\langle g \rangle_k)  \, \int \sqrt{\langle g \rangle_k} \, d^d x \, 
\langle x | e^{T \Delta_{\langle g \rangle_k}} | x \rangle  \nonumber\\
&& \hspace{1.2cm} = 
\frac{  \sum_n   \, \int \sqrt{{\mathbb  S}^{-d}(k, k_0)} \sqrt{\langle g \rangle}_{k_0} \, d^d x \, 
\phi^*_n(x) \phi_{n}(x) \, 
e^{- T \, { \mathbb S}_k (k, k_0) E_n}}{\int \sqrt{{\mathbb  S}^{-d}(k, k_0)} \sqrt{\langle g \rangle}_{k_0}
d^d x}
\nonumber \\
&& \hspace{1.2cm} = 
\frac{\sum_n   e^{- T \, { \mathbb S}_k (k, k_0) E_n}}{\int \sqrt{\langle g \rangle}_{k_0} 
d^d x} \,\,\,\,\,\,  \underrightarrow{{k^2 \approx E_n}} \,\,\,\,\,\,  \sum_n \, \frac{e^{- T \, { \mathbb S} (E_n) E_n}}{V_{\langle g \rangle_{k_0}}}.
\label{proofPT}
\end{eqnarray}
We have used (\ref{kernexp}) and (\ref{opscale0}) from the third to the forth line,
(\ref{norm}) in the last line. 

We assume for a moment 
that $\big<g_{\mu\nu}
\big>_{k_0}$ is an approximately flat metric. In this case the trace in eq.
(\ref{trpropk}) is easily evaluated in a plane wave basis:
\begin{eqnarray}
\label{trplane}
P(T)=\int\frac{d^4p}{(2\pi)^4}\, e^{-p^2\, {\mathbb S}(p)\,T}.
\end{eqnarray}
The dependence from $T$ in (\ref{trplane}) determines the fractal dimensionality of
spacetime via (\ref{specdim}). In the limits $T\rightarrow\infty$ and 
$T\rightarrow 0$ where we are probing very large and small distances,
respectively, we obtain the dimensionalities corresponding to the largest
and smallest length scales possible. The limits $T\rightarrow\infty$ and 
$T\rightarrow 0$ of $P(T)$ are determined by the behavior of ${\mathbb S}(p)$ 
for $p\rightarrow 0$ and
$p\rightarrow\infty$, respectively.

The quantum gravity effects stop below some scale energy 
that we denoted by $k_0$ and 
we have ${\mathbb S}(p\rightarrow 0)=1$. In this case (\ref{trplane}) yields
$P(T)\propto 1/T^2$, and we conclude that the macroscopic spectral dimension
is ${\cal D}_{\rm s}=4$. In the next section we apply the introduced ideas 
to the spatial section in LQG and to the space-time in the covariant spin-foam 
formulation of quantum gravity.


The result we will find about the hight energy spectral dimension 
are of general character. 
The above assumption that $\big<g_{\mu\nu}\big>_{k_0}$ 
is flat was not necessary for obtaining the spectral dimension at 
any fixed scale.  %
This follows from the fact that even for a curved metric the spectral sum
(\ref{trpropk}) can be represented by an Euler-Maclaurin series which always
implies (\ref{proofPT}) as the leading term for $T\rightarrow 0$.\\
{\bf Proof of (\ref{trplane}).}
\begin{eqnarray}
&& P(T) \approx P_{\langle g \rangle_k} (T) = 
V^{-1}(\langle g \rangle_k)  \, \int \sqrt{\langle g \rangle_k} \, d^d x \, 
\langle x | e^{T \Delta_{\langle g \rangle_k}} | x \rangle  \nonumber\\
&& = V^{-1}(\langle g \rangle_k)  \, \int \int \sqrt{\langle g \rangle_k} \, d^d x \, \frac{d^d p}{(2 \pi)^d} \, 
\langle x | e^{T \Delta_{\langle g \rangle_k}} | p \rangle  \langle p | x \rangle  \nonumber\\
&&  = {\mathbb S}(k, k_0)^{d/2} \, V^{-1}(\langle g \rangle_{k_0})  \, \int \int 
{\mathbb S}(k, k_0)^{-d/2}
\sqrt{\langle g \rangle_{k_0}} \, d^d x \, \frac{d^d p}{(2 \pi)^d} \, 
\langle x | p \rangle \, e^{- T \, {\mathbb S} (k, k_0) p^2}  \langle p | x \rangle  \nonumber\\
&& = V^{-1}(\langle g \rangle_{k_0})  \, \int 
\sqrt{\langle g \rangle_{k_0}} \, d^d x \, \int \frac{d^d p}{(2 \pi)^d} \, 
e^{i p x} \, e^{- T \, {\mathbb S} (k, k_0) p^2}  \, e^{- i p x}  \nonumber\\
&& = \int \frac{d^d p}{(2 \pi)^d} \, e^{- T \, {\mathbb S} (k, k_0) p^2}   
\,\,\,\,\,\,  \underrightarrow{{k^2 \approx p^2}} \,\,\,\,\,\,  
\int \frac{d^d p}{(2 \pi)^d} \, e^{- T \, {\mathbb S} (p) p^2}.
\label{dimPOP}
\end{eqnarray}
We have introduced the flat metric $\langle g_{\mu \nu} \rangle_{k_0} = \delta_{\mu \nu}$ and 
$\Delta(k_0) | p \rangle = - p^2 | p \rangle$ in the third line.

%
%


\section{Spectral Dimension in Quantum Gravity}
\renewcommand{\theequation}{5.\arabic{equation}}
\setcounter{equation}{0}
\label{SDQG}
In this section we calculate the spectral dimension of the 
spatial section in LQG and of the space-time for the covariant formulation 
of quantum gravity in terms of spin-foam models \cite{DO}
using the scaling properties introduced in section \ref{SA}, \ref{SVL}. 
I recall here the physical idea explained in the introduction \cite{ASQG}. 
An observer exploring the structure of the spatial section (space-time) using a microscope of resolution 
$l(k)$ ($k$ is the energy scale) will perceive the universe as a Riemannian manifold with the metric 
$\langle g_{ab} \rangle_k$ which is a fixed 
metric at every fixed scale $k$, the quantum space (space-time) can have fractal properties because on different scales different metrics apply. 
In this sense the metric structure on the quantum space (space-time) is given by an infinite set 
$\{ \langle g_{ab} \rangle_k; 0 \leqslant k < +\infty\}$ of ordinary metrics labelled by $k$. 
LQG and Spin-Foams take part in (\ref{specdim}) in the metric 
scaling extrapolated in the sections  \ref{SA}. 

\subsection{Spectral Dimension of the Spatial Section}
We suppose we have a 
Riemannian metric at any energy scale $k$ that we 
denote by $\langle g_{ab} \rangle_k$ ($a,b = 1,2,3$) 
and we probe the space at any scale $0 \lesssim k < +\infty$.
As explained in the previous section, we have to study the properties of the
Laplacian operator of a $3d$ manifold.
Given the scaling properties of the inverse metric (\ref{Fprime}) 
we can deduce the scaling properties of the Laplacian,
\begin{eqnarray}
\Delta(k)= {\mathcal F}(k) \Delta(k_0).
\label{laplacianscale}
\end{eqnarray}
We suppose that the diffusion process involves only a small interval of scales where ${\mathcal F}(k)$
does not change so much. Under this assumption the heat-equation must contain 
$\Delta(k)$ for the specific fixed value of $k$ as explained in the previous section,
\begin{eqnarray}
\partial_{T} K(x, x^{\prime};T) = \Delta(k) K(x, x^{\prime};T),
\label{heatKk}
\end{eqnarray}
If $\Delta(k_0)$ corresponds to flat space,  
the eigenfunctions are plane waves, $\phi_n \rightarrow \phi_p \propto \exp(i p x)$, and 
the eigenvalues of $\Delta(k_0)$ are $- p^2$.
The eigenfunctions resolve length scales $l \approx 1/p$. 
This suggests that when the manifold is probed with a mode of eigenvalue $p^2$, it feels 
the metric $\langle g_{ab} \rangle_k$ for the scale $k = p$. 
The trace of $K(x, x^{\prime};T)$ in the plane wave basis and identifying $k=p$ is
\begin{eqnarray}
&& P(T) 
=  \int \frac{d^3 p}{(2 \pi)^d}  \, 
 {\rm e}^{- T   {\mathcal F}(p) p^2},
\label{PTWP0}
\end{eqnarray}

\begin{figure}
 \begin{center}
\vspace{0.1cm}
  \includegraphics[height=6cm]{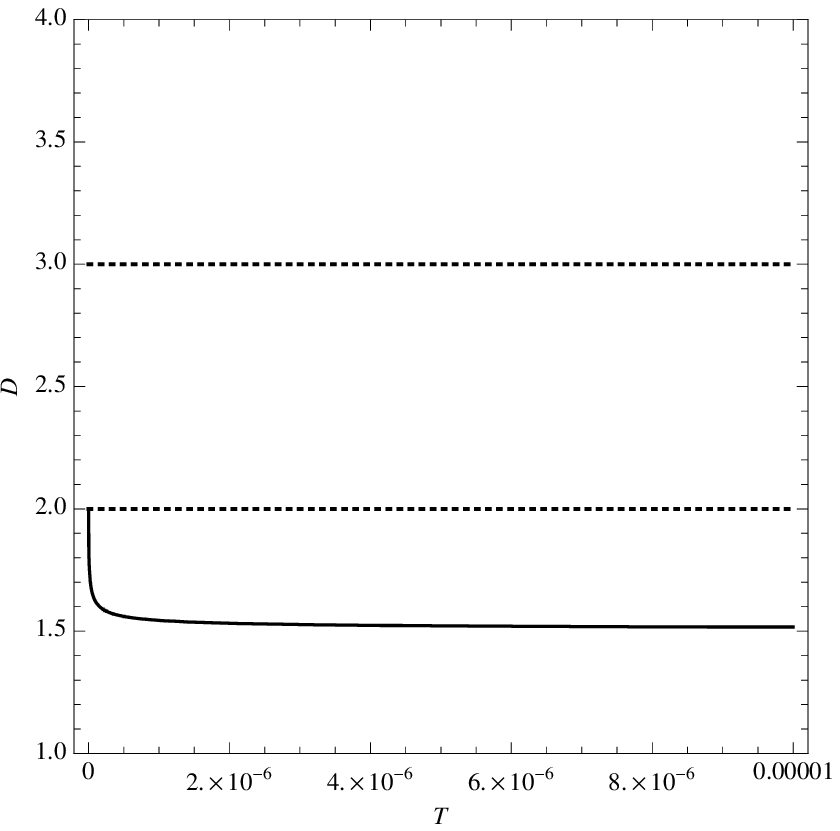}
  \hspace{0.5cm}
  \includegraphics[height=6cm]{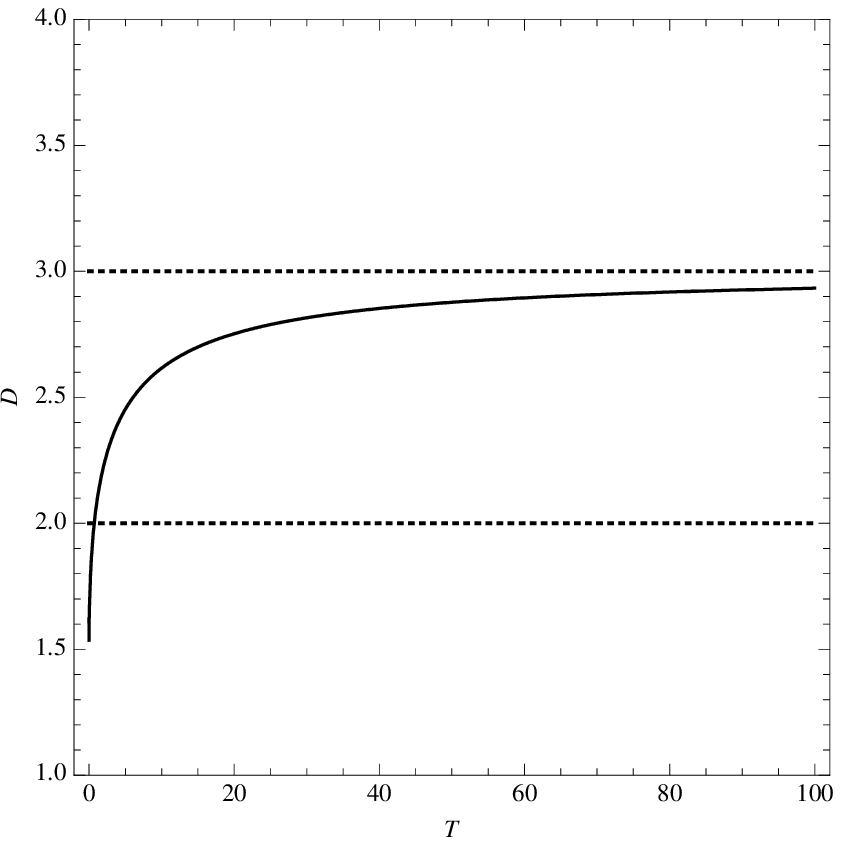}
    \end{center}
    \vspace{-0.1cm}
  \caption{\label{Plot1} 
 Plot of the spectral dimension as function of the diffusion time $T$.
 We can see three different phase from the left to the right as explained in the text.
 }
  \end{figure}

 We have now all the ingredients to calculate the spectral dimension in LQG.
Using the relation (\ref{PTWP0}) and the definition of spectral dimension (\ref{specdim})
we have 
\begin{eqnarray}
{\mathcal D}_s = 2 \, T \frac{\int d^3 p \, {\rm e}^{- p^2 {\mathcal F}(p) T} \, p^2\, 
{\mathcal F}(p)}{\int d^3 p \, {\rm e}^{- p^2 {\mathcal F}(p) T} }.
\label{DSLQG}
\end{eqnarray}
Given the explicit form of the scaling function ${\mathcal F}(k)|_{k=p}$, we are not able to calculate an
analytical solution. We have the spectral dimension (\ref{DSLQG}) 
numerically and obtained the function of $T$ which is plotted in Fig.\ref{Plot1}.
By examining the plot in Fig.\ref{Plot1}, and also calculating analytically the spectral dimension
in the three different regimes of (\ref{Flimits}), we have that
\begin{eqnarray}
{\mathcal D}_s = \left\{ \begin{array}{lll} 
         2 & {\rm for} \,\,\, k \gg E_P, \\ 
         1.5 & {\rm for} \,\,\, k_0 \ll k \ll E_P,\\
        3 & {\rm for} \,\,\, k\gtrsim k_0.
        \end{array} \right. 
\label{regimes}
\end{eqnarray}
We conclude that in LQG, in the case we can consider also the Trans-Planckian regime, 
we have three different phases 
that we will try to interpret in the discussion section. 

We can calculate the spectral dimension of the spatial section using the expectation value of
the metric on the Gaussian states as  explained in section \ref{SVL}. The result is a spectral 
dimension ${\mathcal D}_s^G$ ($G$ denotes Gaussian) that grows from $1.5$ at high energy 
($k \lesssim E_P$) to a low energy value equal to $3$,
\begin{eqnarray}
{\mathcal D}_s^G = \left\{ \begin{array}{lll} 
         1.5 & {\rm for} \,\,\,  k \lesssim E_P,\\
        3 & {\rm for} \,\,\, k\gtrsim k_0.
        \end{array} \right. 
\label{regimesc}
\end{eqnarray}
The plot on the right in Fig.\ref{Plot1} gives the behavior of the spectral dimension 
for Gaussian states.

\subsection{Spectral Dimension of the Space-Time}
\begin{figure}
\vspace{0.15cm}
 \begin{center}
 \includegraphics[height=6.5cm]{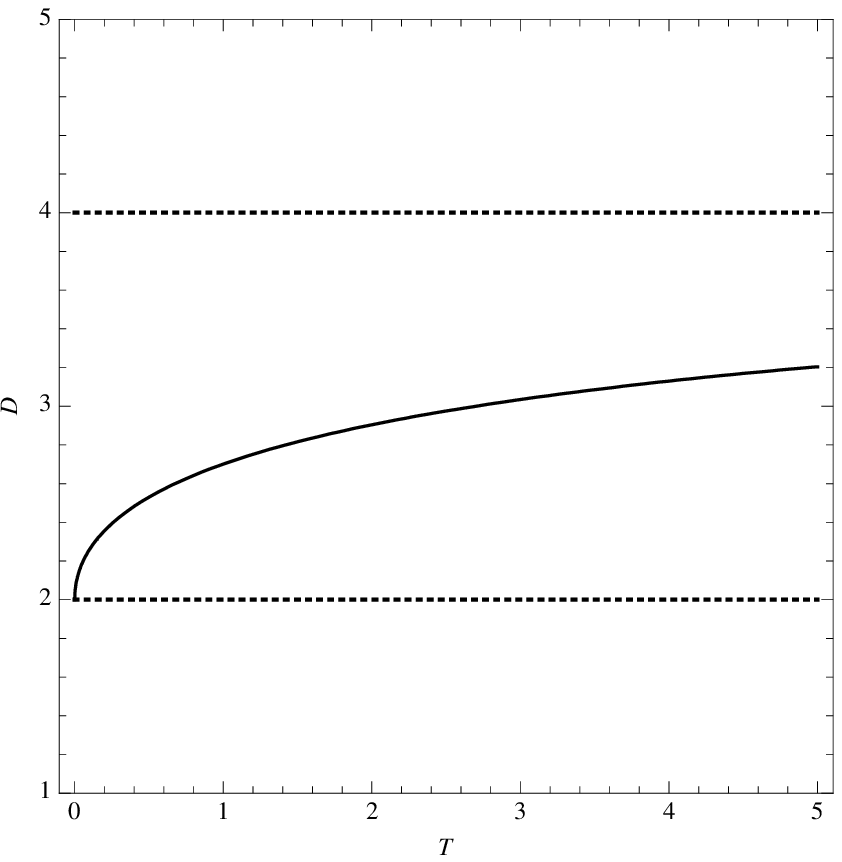}
 \hspace{1.0cm}
 \includegraphics[height=6.5cm]{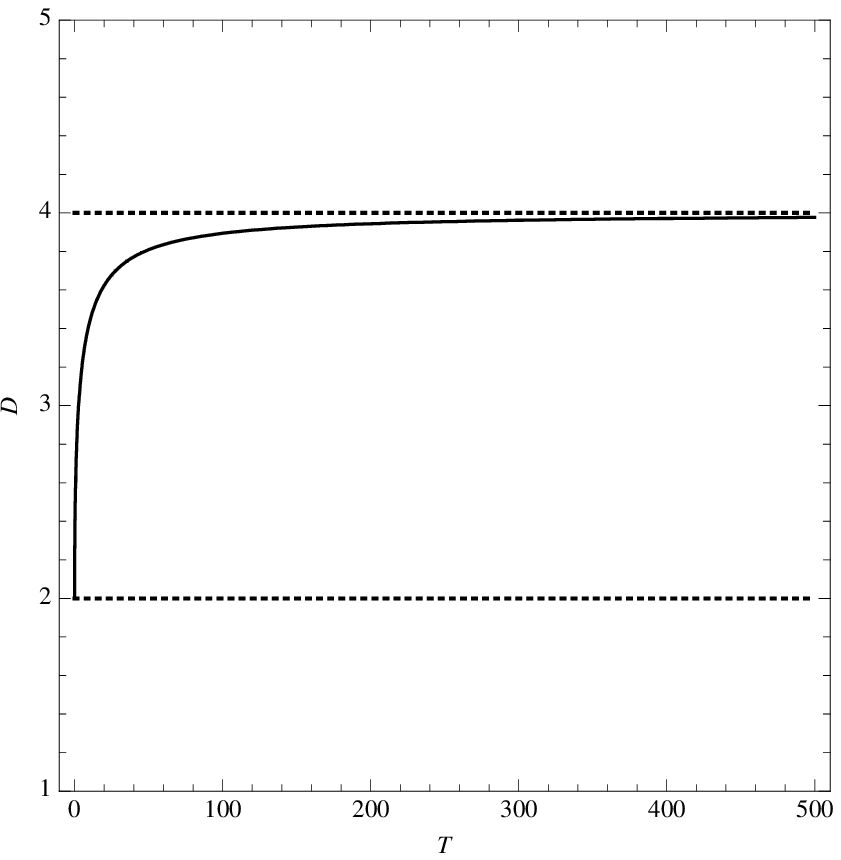}
    \end{center}
  \caption{\label{Plot4} 
Plot of the space-time spectral dimension ${\mathbb D}_s$. We have 
an high energy phase of spectral dimension  ${\mathbb D}_s=2$ and 
a the $4d$ low energy dimension. 
 }
  \end{figure}

In this section we calculate the spectral dimension of the 
Euclidean space-time. The quantity to calculate is the expectation 
value of the operator $\hat{P}_g(T)$,
\begin{eqnarray}
P(T):= \langle \hat{P}_g(T) \rangle_{Phys} = \langle s|  \hat{P}_g(T) | s \rangle_{Phys}
\approx P_{ \langle s|  \hat{g}| s \rangle_{Phys}}(T).
\label{PSF}
\end{eqnarray}
Where in the last approximation we used that $\langle s|  \hat{g}| s \rangle_{Phys}$ is a
dominant 
stationary point in the path integral formulation of the spin-foam model and the scalar 
product is the physical scalar product. It is not necessary to know the stationary point
but it is necessary and sufficient the existence of such a point.
In section (\ref{SFS})
we extracted 
the metric expectation value, and in particular the scaling of the metric expectation value
from the area spectrum, then we obtain 
\begin{eqnarray}
P(T)
\approx P_{ \langle s|  \hat{g}| s \rangle_{Phys}}(T) = 
P\,_{ {\mathbb S}(\ell , \ell_0)\langle s_0|  \hat{g}| s_0 \rangle_{Phys}}(T).
\label{PSF2}
\end{eqnarray}
Using the scaling property of the space-time metric extracted 
in (\ref{SFS}), we are now ready to calculate the
spectral dimension of the $4d$-manifold.
We consider the three possible scaling function introduced in 
section \ref{SFS}, ${\mathbb S}_1$, ${\mathbb S}_2$, ${\mathbb S}_3$ .
We use the notation $\mathbb{D}_{s_i}$ ($i=1,2,3$) for the space-time 
spectral dimension. The quantity to calculate to obtain the spectral dimension is 
\begin{eqnarray}
 {\mathbb D}_{s_i} = 2 \, T \frac{\int d^4 p \, \exp(- p^2 {\mathbb S}_i(p) T) 
 \, p^2\, {\mathbb S}_i(p)}{\int d^4 p \, \exp(- p^2 {\mathbb S}_i(p) T)} \,\,\, , \,\,\,\,\,\, i=1,2,3.
  \label{DSLQG4}
 \end{eqnarray}
\paragraph{Scaling ${\mathbb S}_1$. }
We start considering the scaling function ${\mathbb S}_1(k,k_0) = k^2/k_0^2 +1$
corresponding to the area spectrum $A_j = 2 j l_P^2$ and we identify $k=p$
as explained in the general section on spectral dimension.
We calculate numerically (\ref{DSLQG4}) and we plot the result in
Fig.\ref{Plot4} 
as a function 
 of the diffusion time $T$. For $T\rightarrow 0$ (or $k\rightarrow \infty$) we 
 obtain the spectral dimension ${\mathbb D}_s =2$ and for $T \rightarrow \infty$ (or $k \rightarrow 0$)
 we obtain ${\mathbb D}_{s} =4$.
 We consider the high and low energy limits obtaining the following
 behavior of the spectral dimension,
 \begin{eqnarray}
{\mathbb D}_{s_1} = \left\{ \begin{array}{lll} 
         2 & {\rm for} \,\,\, k  \gtrsim E_P, \\ 
                 4 & {\rm for} \,\,\, k \ll E_P.
        \end{array} \right. 
\label{DSLQG4dlimit}
\end{eqnarray}
This result in space-time is in perfect accord 
 with the results in CDT \& ASQG \cite{CDT}, \cite{ASQG}.
 
 \paragraph{Scaling ${\mathbb S}_2$. }
 If we use the scaling function ${\mathbb S}_2(k, k_0)$ defined at the end of section \ref{SFS},
 we obtain the same behavior for the spectral dimension in the case in which we
 consider the ultraviolet cutoff $k < E_P$.
 If we consider the regime in which the momentum $k \geqslant E_P$, 
 we obtain the spectral dimension ${\mathbb D}_{s_2} = 4$ as $T\rightarrow 0$ 
 (or $k\rightarrow + \infty)$. The behavior of ${\mathbb D}_s$ is the same
as in (\ref{DSLQG4dlimit}) for $k < E_P$.
 \begin{eqnarray}
{\mathbb D}_{s_2} = \left\{ \begin{array}{lll} 
          4 & {\rm for} \,\,\, k  \gg E_P, \\ 
         2 & {\rm for} \,\,\, k  \lesssim E_P, \\ 
                 4 & {\rm for} \,\,\, k  \gtrsim  k_0.
        \end{array} \right. 
\label{DSLQS2}
\end{eqnarray}

This high energy behavior of the spectral dimension is appealing if
we consider the space-time Ricci invariant ${\rm R}(g) = {\rm R}^{\mu}_{\mu}(g)$.  
Under the rescaling ${\mathbb S}_2(k,k_0)$, the Ricci curvature scales as:
${\rm R}(g)_k \approx {\mathbb S}_2(k, k_0) {\rm R}(g)_{k_0}$. 
At short distances, or $k\rightarrow +\infty$, 
${\rm R}(g)_k$ is upper bounded as can be seen by considering the limit: 
$\lim_{k\rightarrow \infty} {\mathbb S}_2(k,k_0) \propto (E_p/k_0)^2$.
The upper bound of the curvature could be a sign of singularity
problem resolution as shown in cosmology and black holes using the 
minisuperspace simplification of quantum gravity \cite{CBH}.

\paragraph{Scaling ${\mathbb S}_3$. }
We conclude the section by considering the case in which the area spectrum is
$A_j= l_P^2 \sqrt{j(j+1)}$. In this case, the scaling function is 
the same as given in (\ref{Fprime}) but the momentum $k$ is now four dimensional.
The spectral dimension has the same behavior as plotted in
Fig.\ref{Plot4} for the case $k<E_P$, but instead ${\mathbb D}_s=8/3$ in
the trans-planckian limit ($k\gg E_P$). 
However, if we do not consider the trans-Planckian limit, 
we obtain the same spectral dimension (\ref{DSLQG4dlimit})
for any form of the area spectrum considered in this section.
\begin{eqnarray}
{\mathbb D}_{s_3} = \left\{ \begin{array}{lll} 
          8/3 & {\rm for} \,\,\, k  \gg E_P, \\ 
         2 & {\rm for} \,\,\, k  \lesssim E_P, \\ 
                 4 & {\rm for} \,\,\, k  \gtrsim  k_0.
        \end{array} \right. 
\label{DSLQS3}
\end{eqnarray}

\paragraph{Continuum approximation of the representations $j$.}
We want to show here the validity and the limitation of the simplification 
to consider $j$, and then $\ell = l_P \sqrt{j}$, as a continuum variable. 
We recall that the spacing in the $SU(2)$ representation $j$ is $\Delta j = 1/2$
than the distance between two neighbours points is 
$\Delta \ell = l_P(\sqrt{ j + 1/2} - \sqrt{j})$. For large $j$, $\Delta \ell \rightarrow 0$
but for $j=0$, $\Delta \ell = l_P/\sqrt{2}$, 
therefore we can approximate the discrete variable $j$ with a continuum variable   
until the wavelength of a probe field is $\lambda \gtrsim l_P/\sqrt{2}$.
For the momentum $k$ the condition 
is $k \lesssim \sqrt{2} E_P$. 
The analysis in this
section is correct until the Planck scale but we must be carefully 
in the case $j=0$.
In the case $j=0$ we can consider a region (or radius $\ell \ll l_P$) around $j=0$
where the discrete scaling functions (${\mathcal F}, {\mathbb S}_i$) 
are replaced 
by smooth and continuum functions around $j=0$ assuming the same scaling of the 
discrete scaling functions.
Using this definition we can calculate the spectral dimension for $T\ll l_P$ and we obtain 
 results coincident with those anticipated in this section
in the trans-Planckian regime. We want to emphasize 
that for the scaling function ${\mathbb S}_2(k, k_0)$ it is not useful to introduce a 
smooth function around $j=0$ to define the integral, because the area spectrum has a natural minimum 
$A_{\rm Min} = l_P^2( 2 j+1)|_{j =0} = l_P^2$. This means that 
${\mathbb S}_2(k \rightarrow +\infty,k_0) \approx const.$ and the integral (\ref{DSLQG4}) 
gives ${\mathbb D}_{s_2} = 4$ for $k \rightarrow + \infty$.
%
  %
   \begin{figure}
 \begin{center}
 \includegraphics[height=5cm]{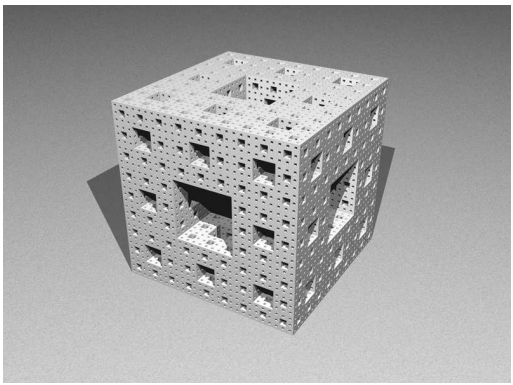}
    \end{center}
    \vspace{-0.3cm}
  \caption{\label{Plot2} 
  This is the Menger sponge of fractal dimension $\log(20)/\log(3) \approx 2.7268$.
  In general we can imagine the spatial section or the space-time to have
  many holes at all the scales. 
  A probe scalar field 
  feels the holes only when the diameter of the hole is comparable with
  its weave length. 
 }
  \end{figure}

\subsection{To Avoid the Singularities}
In this section we want to use the results in \ref{SFS} to show
that the space-time could be intrinsically singularity free
in quantum gravity. 

\begin{figure}
 \begin{center}
 \includegraphics[height=6cm]{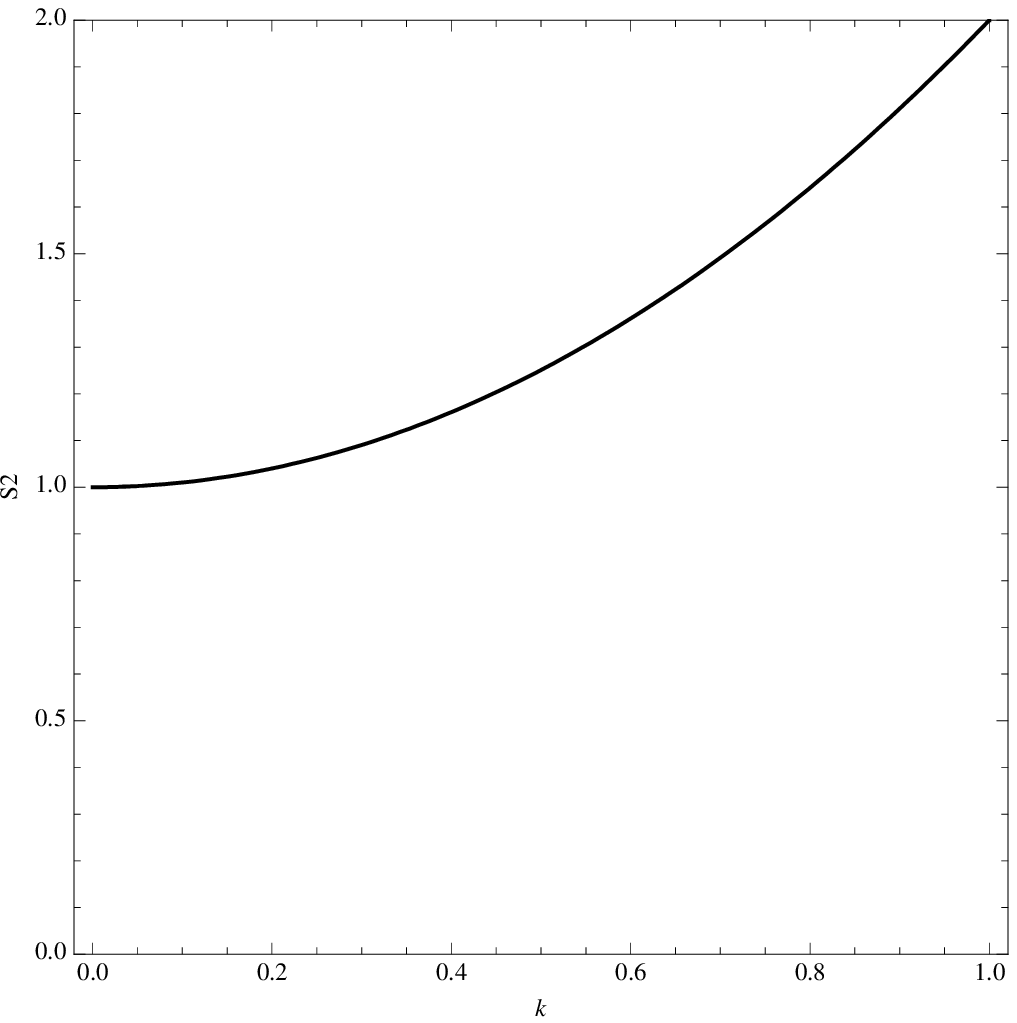}
 \hspace{1cm}
  \includegraphics[height=6cm]{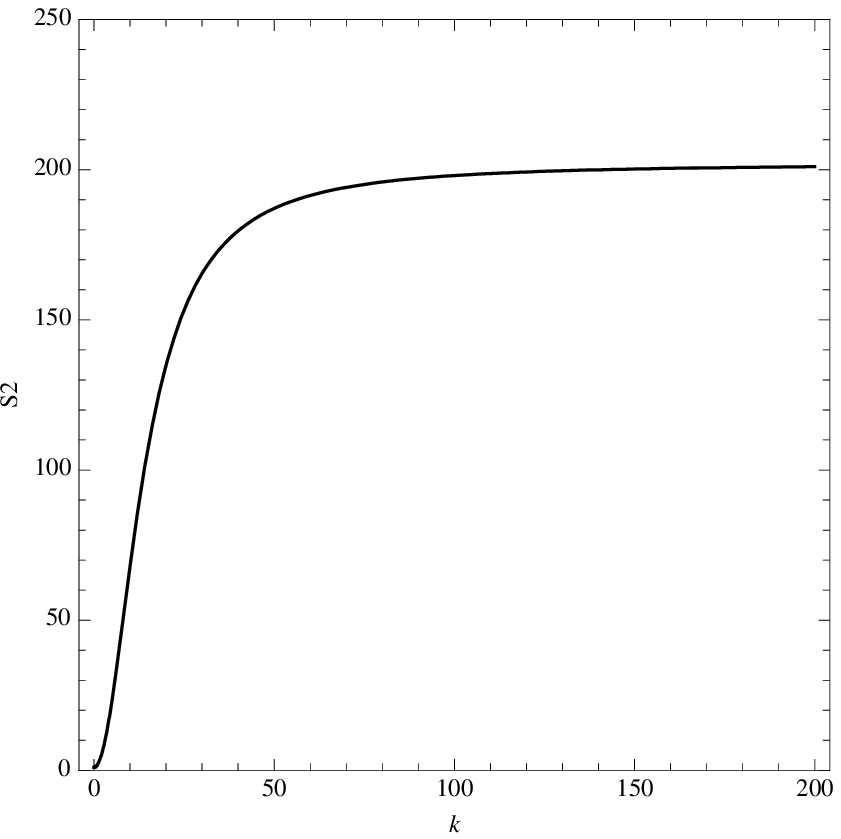}
    \end{center}
    \vspace{-0.3cm}
  \caption{\label{Plot2c} 
  Plot of the scaling function ${\mathbb S}_2(k, k_0)$ for $k_0=1$ to $E_P=10$.
 }
  \end{figure}

We consider all the possible action terms compatible 
with the Diff-invariance and we label them by 
$I_n[g_{\mu\nu}] = \int d^4 x \, {\mathcal L}_n$. We can recall same curvature scalar densities,
\begin{eqnarray}
&& {\mathcal L}_0 = \sqrt{g} \, , \,\,\, {\mathcal L}_1 = \sqrt{g} R, \nonumber \\
&&{\mathcal L}_2^a = \sqrt{g} R^2 \, , \,\,\,{\mathcal L}_2^b = \sqrt{g} R_{\mu \nu}^2, \,\, , \,\,\, 
{\mathcal L}_2^c = \sqrt{g} R_{\mu\nu\rho\sigma}^2 \, , \,\,\, 
{\mathcal L}_2^d = \sqrt{g} \, \nabla_{\mu} \nabla^{\mu} R, \nonumber \\
&& {\mathcal L}_3^a = \sqrt{g} R^3\, , \,\,\, {\mathcal L}_3^c = \sqrt{g} R_{\mu\nu}^3 \, , \,\,\,  I_3^b = \sqrt{g} R_{\mu\nu\rho\sigma}^3 \, ,\,\,\, {\mathcal L}_3^c = \sqrt{g} R_{\mu\nu}^3, \nonumber \\
&& {\mathcal L}_3^d = \sqrt{g} R \, R_{\mu\nu}^2  \, , \,\,\, 
{\mathcal L}_3^e = \sqrt{g} \, \nabla_{\mu} R_{\nu \rho} \, \nabla^{\mu} R^{\nu \rho}
\, , \,\,\, \dots .
\label{curInv}
\end{eqnarray}
The mass dimension of $I_n$ are
indicated by  $-d_n$, $[I_n] = - d_n$. Explicitly 
\begin{eqnarray}
d_n=\{\overbrace{\underbrace{4, 2, 0, -2 , -4, \dots , }}^{1, R, R^2, R^3, R^4 , \dots}_{n=0 , n=1, n=2, n=3, n=4, \dots } \}.
\label{dime}
\end{eqnarray}
Under a scaling of the metric $g_{\mu \nu} \rightarrow c^2 g_{\mu \nu}$, we have 
the correspondent scaling of the action terms 
\begin{eqnarray}
I_n[g_{\mu \nu}] \rightarrow I_n[c^2 g_{\mu \nu}] = c^{d_n} I_n[g_{\mu \nu}]
\label{scaleact}
\end{eqnarray}
Using (\ref{scaleact}) we obtain the scaling of the curvature invariant $C_n:={\mathcal L}_n/\sqrt{g}$,
\begin{eqnarray}
C_n[ c^2 g_{\mu \nu}] \rightarrow C_n[c^2 g_{\mu \nu}] = c^{d_n - 4} C_n[g_{\mu \nu}].
\label{scalcurv}
\end{eqnarray}
In quantum gravity we consider the expectation value of the invariants and 
using the notation of section \ref{SFS} we obtain
\begin{eqnarray}
\frac{\langle C_n[  g_{\mu \nu}] \rangle_k}
{\langle C_n[ g_{\mu \nu}] \rangle_{k_0}} \approx 
\frac{ C_n[  \langle g_{\mu \nu} \rangle_k ] }
{C_n[  \langle g_{\mu \nu} \rangle_{k_0}]} =
\frac{ C_n[ {\mathbb S}^{-1}_i(k, k_0) \langle g_{\mu \nu} \rangle_{k_0}]}
{C_n[  \langle g_{\mu \nu} \rangle_{k_0}]} = {\mathbb S}_i(k, k_0)^{\frac{4-d_n }{2}}.
\label{scalcurv2}
\end{eqnarray}
If the scaling function ${\mathbb S}_i(k, k_0)$ is upper bounded 
all the curvature invariant $C_n$ are upper bounded.
This is the case when the scaling function is ${\mathbb S}_2$. 
We can consider for example of $C_1 = R$: 
\begin{eqnarray}
\frac{ \langle R[ g_{\mu \nu}] \rangle_k } 
{\langle R[ g_{\mu \nu} ] \rangle_{k_0} } \approx \frac{ R[ {\mathbb S}^{-1}_2(k, k_0) \langle g_{\mu \nu} \rangle_{k_0}]}
{R[ \langle g_{\mu \nu} \rangle_{k_0}]} = {\mathbb S}_2(k, k_0).
\label{R}
\end{eqnarray}
The curvature $R$ is regular at any energy scale because 
the scaling function is limited from ${\mathbb S}_2 \approx 1$ when $k \lesssim k_0$ 
and ${\mathbb S}_2 \approx 2 E_P^2/k_0^2$ for $k \gg E_P$ as is represented in Fig.\ref{Plot2c}. 
This result is strongly related to the area spectrum $A_j = l_P^2 (2 j +1)$. 

\begin{figure}
 \begin{center}
  \includegraphics[height=11.0cm]{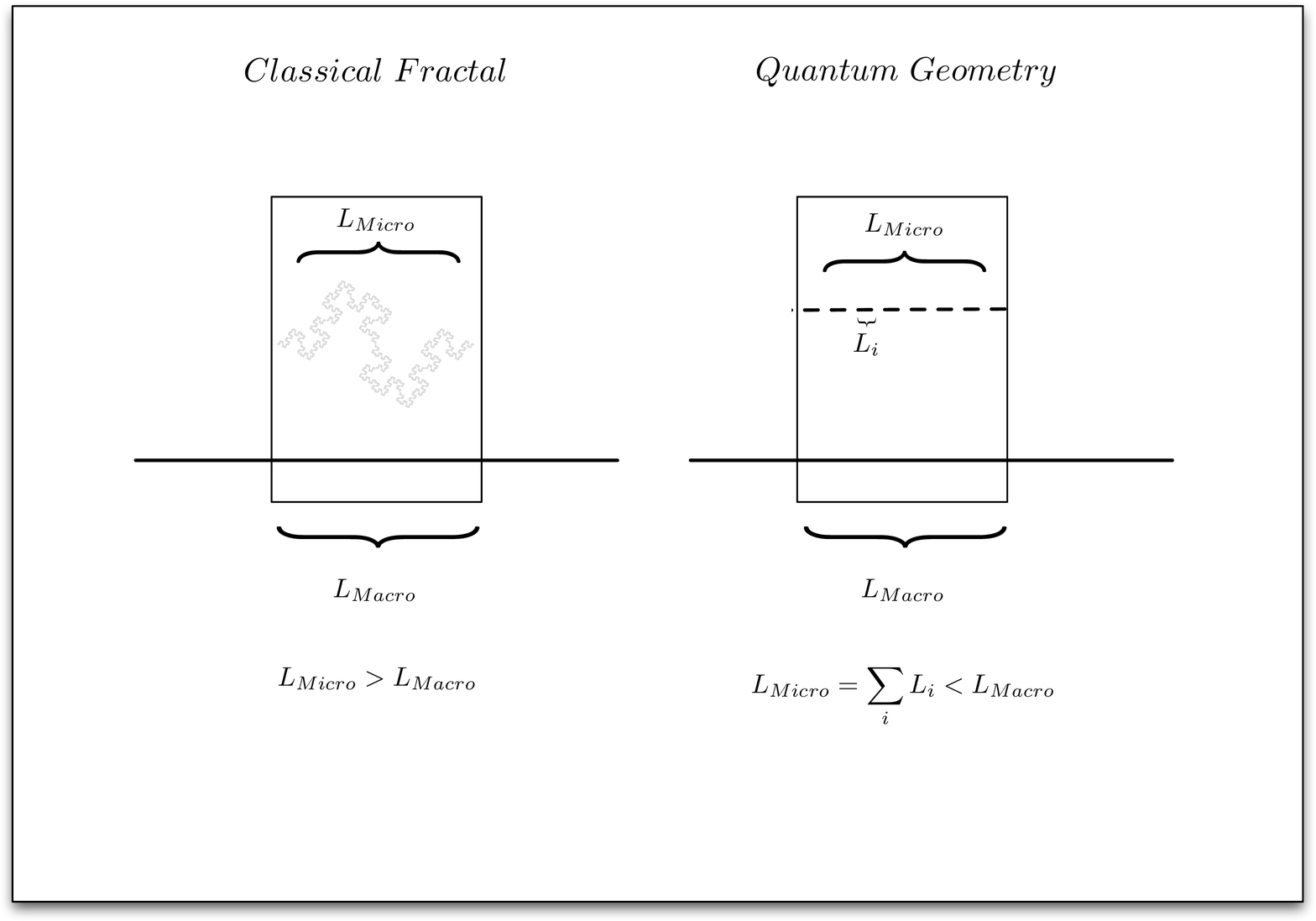}
    \end{center}
    \vspace{-0.3cm}
  \caption{\label{FC} 
In this picture we represent schematically the difference and similarities between  
 a one dimensional fractal and the fractal structure of quantum space or space-time.
 The result of our analysis is that the spatial section and the space-time 
 show a lower spectral dimension at hight energy. This result can be understand 
 recalling the scaling of the metric. In quantum gravity (in all the cases studied and 
 until the Planck scale)
 the length becomes smaller when we increase the energy of a probe scalar field
 and this is represented for 
 the one dimensional case in the picture on the right. 
 We interpret the result in the following way. At high energy the spectral dimension
is ${\mathcal D}_s < 4$ (or the space-time) or ${\mathcal D}_s < 3$ (for the spatial section) 
resembles a manifold which presents holes. 
At the Planck scale the manifold seems to present a large number of holes. 
 We can compare the result in quantum geometry with a simple fractal.
 For a one dimensional fractal like the one in the
 picture on the left the distance at hight energy is bigger then at low energy.
 The behavior of quantum geometry is the contrary.
 } 
  \end{figure}

\section{Conclusions and Discussion}  
\renewcommand{\theequation}{6.\arabic{equation}}
\setcounter{equation}{0}
In this paper we calculated explicitly the spectral dimension (${\mathcal D}_s$)
of the {\bf spatial section} in LQG using: (i) the area spectrum scaling on spin-network states, 
(ii) the scaling of the volume 
and length operators on Gaussian states. The result is the same and 
differences show up only in the trans-Planckian regime.
We obtained ${\mathcal D}_s$ as a function
of a fictitious time $T$ needed for a probe scalar field
to diffuse in the manifold or equivalently as a function of diff-invariant length scale.
In both cases (i) and (ii) we have the same behavior from $1.5$ at high energy 
to $3$ at low energy. In the case (i) if we boost the momentum 
beyond the Planck energy we have three phases: a short scale phase $l \ll l_P$ of spectral 
dimension ${\mathcal D}_s = 2$, an intermediate scale phase $l_P \ll l \ll l_0$ of spectral 
dimension ${\mathcal D}_s = 1.5$ and a large scale phase with ${\mathcal D}_s = 3$
($l_0$ represent the infrared large scale).

We calculated moreover the spectral dimension for the {\bf space-time} in
the spin-foam models framework using the scaling of the area operator 
in three different cases. 
In the first case we obtained ${\mathbb D}_s =2$
at the Planck scale and ${\mathbb D}_s =4$ at low energy.
This result is the same as obtained in CDT \& ASQG \cite{CDT}, \cite{ASQG}.
A different area spectrum (the other two cases)
 that comes from a different quantum ordering 
\cite{DO} gives the same result until the Planck
scale but a new different behavior in the trans-Planckian regime. 
We can interpret the result in the following way.
First of all, we want to underline that the probe scalar field we used is just 
a fictitious field and not a physical scalar field. 
The metric scales as 
$\langle g_{\mu \nu} \rangle_k \approx {\mathbb S}_i \langle g_{\mu \nu} \rangle_{k_0}$,
 then when we increase the energy of the scalar field applying the scalings 
 ${\mathbb S}_1$ and ${\mathbb S}_3$ it sees a smaller and smaller distance 
 until $\ell \approx 0$, but when we applying ${\mathbb S}_2$ the smaller 
 microstructure it is able to see is the Planck scale.
The space-time is a sort of sponge with many holes \cite{SC} 
that the field is able to feel only increasing the energy.

We conclude comparing the our result  to the
spectral dimensions of the spacetime which were obtained by Monte
Carlo simulations of the causal dynamical triangulation model \cite{CDT}:
\begin{eqnarray}
\label{mcspecdims}
{\cal D}_{\rm s}(T\rightarrow\infty)&=&4.02\pm 0.1\nonumber\\
{\cal D}_{\rm s}(T\rightarrow 0)&=&1.80\pm 0.25
\end{eqnarray}
These suggest that the long-distance and short-distance spectral
dimension should be 4 and 2, respectively. This result coincides 
with our in the space-time case.
The space-time result in this paper is supported by an explicit analysis of the dynamics 
in $3d$-spinfoam models \cite{FL}.  



{\em Acknowledgements.} 
We are extremely grateful to  Dario Benedetti, Daniele Oriti, Eugenio Bianchi, Tim Koslowski,
James Ryan, 
and Francesco Caravelli.

\end{document}